\documentclass{aa}
\usepackage{natbib}
\usepackage{graphicx}
\usepackage{txfonts}

\begin{document}
\title{
Characterization of the HD~17156 planetary system
\thanks{
Based on observations made with the Italian Telescopio Nazionale Galileo
(TNG) operated on the island of La Palma by the Fundacion Galileo Galilei
of the INAF (Istituto Nazionale di Astrofisica) at the Spanish Observatorio
del Roque de los Muchachos of the Instituto de Astrofisica de Canarias.
Based on observations collected at Asiago observatory, at Observatoire
de Haute Provence and with Telast at IAC.
}
}

\author{
M. Barbieri               \inst{1}    \and
R. Alonso                 \inst{1}    \and
S. Desidera               \inst{2}    \and
A. Sozzetti               \inst{3}    \and
A.F. Martinez Fiorenzano  \inst{4}    \and
J. M. Almenara            \inst{5}    \and
M. Cecconi                \inst{4}    \and
R.U. Claudi               \inst{2}    \and
D. Charbonneau            \inst{7}    \and
M. Endl                   \inst{8}    \and
V. Granata                \inst{2,9}    \and
R. Gratton                \inst{2}    \and
G. Laughlin,              \inst{10}    \and
B. Loeillet               \inst{1,6}
and Exoplanet Amateur Consortium
\thanks{
E.A.C. observations obtained by:
F. Castellani (Mt. Baldo Observatory),
B. Gary,
J. Gregorio,
C. Lopresti,
A. Marchini (Siena University Observatory),
M. Nicolini (Cavezzo Observatory),
R. Papini,
C. Vallerani
}
}

\authorrunning{M. Barbieri et al.}
\offprints{M. Barbieri, \email{mauro.barbieri@oamp.fr} }
\institute{
Laboratoire d'Astrophysique de Marseille, 38 rue Joliot-Curie, 13388 Marseille Cedex 13, France                           \and
INAF Osservatorio Astronomico di Padova, Vicolo dell' Osservatorio 5, I-35122, Padova, Italy                              \and
INAF Osservatorio Astronomico di Torino, 10025 Pino Torinese, Italy                                                       \and
Fundaci\'on Galileo Galilei - INAF, Rambla Jos\'e Ana Fern\'andez P\'erez 7, 38712 Bre\~na Baja (TF), Spain               \and
Instituto de Astrof\'\i sica de Canarias, C/V\'\i a L\'actea s/n, E-38200, La Laguna, Spain                               \and
IAP, 98bis Bd Arago, 75014 Paris, France                                                                                  \and
Harvard-Smithsonian Center for Astrophysics, 60 Garden Street, Cambridge, MA 02138, USA                                   \and
McDonald Observatory, The University of Texas at Austin, Austin, TX 78712, USA                                            \and
CISAS, Universit\`a di Padova                                                                                             \and
University of California Observatories, University of California at Santa Cruz Santa Cruz, CA 95064, USA
}

\date{ Received December 03 2008 / }

\abstract
{}
{
To improve the parameters of the HD~17156 system
(peculiar due to the eccentric and long orbital period of its transiting planet)
and constrain the presence of stellar companions.
}
{
Photometric data were acquired for 4 transits, and high precision radial velocity
measurements were simultaneously acquired with SARG@TNG for one transit.
The template spectra of HD~17156 was used to derive effective temperature, gravity, and metallicity.
A fit of the photometric and spectroscopic data was performed to measure the
stellar and planetary radii, and the spin-orbit alignment.
Planet orbital elements and ephemeris were derived from the fit.
Near infrared adaptive optic images was acquired with ADOPT@TNG.
}
{
We have found that the star has a radius of R$_S = 1.44\pm 0.03$R$_\odot$
and the planet R$_P = 1.02\pm 0.08$R$_J$.
The transit ephemeris is T$_c = 2\,454\,756.73134 \pm 0.00020 + N\cdot 21.21663 \pm 0.00045$ BJD.
The analysis of the Rossiter-Mclaughlin effect shows that the system is spin orbit
aligned with an angle $\lambda = 4.8^\circ\pm 5.3^\circ$.
The analysis of high resolution images has not revealed any stellar companion with
projected separation between 150 and 1\,000 AU from HD~17156.
}
{}

\keywords{
stars: individual: HD 17156 --  binaries: eclipsing -- planetary systems --
techniques: spectroscopic, photometric
}

\maketitle
\newcommand{\hd}{{HD~17156}~}
\newcommand{\hdb}{{HD~17156b}~}

\section{Introduction}

The discovery of transiting extrasolar planets (TESP)
is of special relevance
for the study and characterization of planetary systems.
The combination of photometric and radial velocity measurements
allows to measure directly the mass and radius of an exoplanet, and hence its
density, which is the primary constraint on a planet's bulk composition.
Dedicated follow-up observations of TESP during primary transit and secondary eclipse at
visible as well as infrared wavelengths allow direct measurements of planetary
emission and absorption (e.g., \citet{2007prpl.conf..701C}, and references therein).
Transmission spectroscopy during primary
eclipse in recent years has been successful in characterizing the atmospheric
chemistry of several Hot Jupiters (e.g., \citet{2002ApJ...568..377C} and \citet{2007Natur.448..169T}).
Infrared measurements gathered at a variety of orbital phases, including
secondary eclipse, have permitted the
characterization of the longitudinal temperature profiles of nearby TESP.
The quickly increasing amount of
high-quality data obtained for TESP has provided the first crucial constraints
on theoretical models describing the physical structure and the atmospheres of gas
and ice exoplanets.
The detailed characterization of TESP ultimately is of special relevance to test several proposed
formation and orbital evolution mechanisms of close-in planets.

The planet \object{HD~17156b}, detected by \cite{2007ApJ...669.1336F}(hereinafter F07)
using the radial velocity method, was  shown to transit in
front of its parent star by \cite{2007A&A...476L..13B}. Additional
photometric measurements were presented in follow-up papers by
\cite{2008A&A...485..871G}, \cite{2008PASJ...60L...1N},
\cite{2008ApJ...681..636I}, \cite{2008arXiv0810.4725W}.
This planet is unique among the known transiting
systems in that its period (21.2 days) is more than 5 times longer than
the average period for this sample, and it has the largest eccentricity
($e = 0.67$).

\cite{1910PAllO...1..123S}, \cite{1924ApJ....60...15R} and \cite{1924ApJ....60...22M}
showed that a transiting object, like a stellar companion, produces a distortion
in the stellar line profiles due to the partial eclipse of the rotating stellar
surface during the event, and thus an apparent anomaly in the measured
radial velocity of the observed primary star. In the case of close-in giant planets
transiting solar-type stars, the amplitude of the Rossiter-McLaughlin
effect ranges typically between a few and $\sim100$ m s$^{-1}$,
depending on orbital period, stellar and planetary radii, and
stellar rotational velocity. This is within the current instrument
capabilities for the brightest transiting planets. Indeed, this effect
has been previously observed on several TESP
(e.g. \citet{2008arXiv0807.4929W} and references therein).
The observation of the Rossiter-McLaughlin effect allows to measure
the relative inclination angle between the sky projections of the orbital plane
and the stellar spin axis.
Another important result derived from these observations is the possibility
to determine whether short orbital period transiting planets move in the same
direction of the stellar spin, indicating the existence of strong dynamical interactions
between the parent star and its planet. Measurements of the Rossiter-McLaughlin
effect thus provide relevant fossil evidence of planet formation and migration
processes, as well as dynamical interactions with perturbing bodies
\citep{2002Icar..156..570M}

Up to now 8 of the 9 transiting planets for which the Rossiter-McLaughlin effect was measured
are coplanar systems, as our Solar System (the
orbital axes of all the planets and the Sun spin axis are aligned
within a few degrees). This is compatible with a formation mechanism
for close-in giant planets including migration via tidal interactions
with the protoplanetary disk. Only the XO-3 planet seems to be a non aligned
system \citep{2008A&A...488..763H}.

The measurement of the Rossiter-McLaughlin effect is of special
relevance for the \object{HD~17156} system, given the very high eccentricity
and orbital period, much longer than the other transiting planets.
With a period of 21 days, \object{HD~17156b} is well outside the
peak in the period distribution of close-in planets at about 3 days,
possibly implying a different migration history with respect to the
other transiting planets. An eccentricity as high as that of
\object{HD~17156b} can be hardly explained by models of planet migration
via tidal interactions with the protoplanetary disk.
Alternatively, high-eccentricity planets might be the outcome of a variety of
possible planet-planet dynamical interactions \citep{2002Icar..156..570M}.
In these scenarios, high relative inclinations between the stellar rotation axis and
the planet orbit plane are possible.
An additional way to get large relative inclinations and eccentricities is
represented by Kozai resonances with close stellar companion.
Interestingly, a rather large planetary mass coupled with a short orbital period might
be an indication for binarity, as typically high-mass short-period planets
occur in binary systems \citep{2007A&A...462..345D}.

\cite{2008PASJ...60L...1N} presented the first observations of the
Rossiter-McLaughlin effect in \object{HD~17156}. They found an angle
between the sky projections of the orbital axis and the stellar rotation
axis $\lambda=62\pm25^{\circ}$. However, \cite{2008ApJ...683L..59C}
did not confirm this claim, suggesting instead well-aligned axes.
Such a discrepancy calls for additional high-precision RV monitoring
during planetary transit. These are presented in this work, along with
additional photometric observations and a critical revision of
stellar and planetary parameters.
To further constrain the origin of the special properties of \hdb,
we also searched for possible wide stellar companions using adaptive optics observations.

The overall organization of this paper begins in Sec.~\ref{s:hrs}
with the description of high resolution spectroscopy data obtained with SARG@TNG.
The following Sec.\ref{s:star} covers the analysis of the stellar parameters.
Sec.\ref{s:photom} presents the photometric data collected during several planetary transits.
Next Sec.\ref{s:system} describes the analysis of the radial velocity and photometric
data, and their results. In Sec.\ref{s:binarity} we present the results for the
search of additional stellar companions to \hd.
Finally in Sec.\ref{s:concl} our conclusion are presented.

\section{High resolution spectroscopy}
\label{s:hrs}

We observed \hd on 2007 December 3, including continuous monitoring
(about 8 hours) during the transit, with SARG, the high resolution
spectrograph of the TNG \citep{2001ExA....12..107G}. Observing
conditions were not optimal, with seeing ranging from 1.3\arcsec to 2.0
\arcsec.
We obtained the stellar template (without the iodine cell) first and then started the
uninterrupted series of observations with the iodine cell.
Exposure time of the spectra
acquired with the iodine cell was 900~s, typically resulting in S/N of
about 80 per pixel.
One additional spectrum was obtained on 25 Oct 2007.
These spectra were reduced and analyzed as those for the ongoing
planet search program with SARG \citep{2007arXiv0705.3141D} using the
AUSTRAL code \citep{2000A&A...362..585E}. Table~\ref{t:rv} lists the
radial velocities.

\begin{table}
\centering
\caption{Differential radial velocities and bisector velocity span
of \hd obtained with SARG at TNG.}
\label{t:rv}
\begin{tabular}{rrrrrrrrr}
\hline
\hline
BJD            &  RV         &  err    &    span  & err    \\
               &  m/s        &  m/s    &    m/s   & m/s    \\
\hline
 2454398.58856   & -81.03  &    5.76    &          &        \\
 2454438.40203   & 100.47  &    4.55    &  -25.47  & 63.07  \\
 2454438.41331   &  82.40  &    4.56    &  -32.78  & 62.90  \\
 2454438.42460   &  76.72  &    4.55    &   19.04  & 60.32  \\
 2454438.43608   &  90.10  &    4.43    &   74.70  & 59.30  \\
 2454438.44737   &  91.46  &    3.98    &   -5.58  & 56.75  \\
 2454438.45866   &  76.28  &    4.35    &   64.69  & 59.89  \\
 2454438.47011   &  65.86  &    3.98    &   47.68  & 57.10  \\
 2454438.48140   &  60.08  &    4.01    &   40.49  & 53.96  \\
 2454438.49270   &  30.74  &    3.72    &  119.82  & 57.44  \\
 2454438.50415   &  32.97  &    3.98    &  216.98  & 59.74  \\
 2454438.51544   &  20.63  &    4.49    &  108.08  & 57.92  \\
 2454438.52671   &  18.78  &    4.59    &   74.54  & 57.66  \\
 2454438.53816   &   4.29  &    4.29    &   82.66  & 58.07  \\
 2454438.54946   &  -3.77  &    4.13    &   90.90  & 57.79  \\
 2454438.56075   &   7.21  &    4.93    &   41.59  & 63.36  \\
 2454438.57218   &  -3.84  &    3.77    &   58.41  & 57.50  \\
 2454438.58349   & -11.52  &    4.53    &   50.75  & 60.76  \\
 2454438.59478   & -31.71  &    4.43    &   45.47  & 59.96  \\
 2454438.62486   & -26.61  &    4.74    &   36.70  & 64.19  \\
 2454438.63615   & -43.01  &    4.01    &   18.82  & 58.87  \\
 2454438.64745   & -46.79  &    4.23    &   66.00  & 65.44  \\
 2454438.65894   & -54.19  &    4.77    &  -38.79  & 65.29  \\
 2454438.67023   & -55.35  &    4.72    &    3.19  & 67.30  \\
 2454438.68152   & -64.85  &    5.32    &  102.93  & 70.23  \\
 2454438.69306   & -71.94  &    5.20    &   73.98  & 67.86  \\
 2454438.70435   & -82.48  &    5.08    &  111.26  & 72.47  \\
 2454438.71584   & -88.29  &    5.13    &  -18.92  & 72.31  \\
 2454438.72713   & -92.59  &    5.15    &   94.27  & 72.55  \\
\hline
\end{tabular}
\end{table}

The absolute radial velocity of \hd was derived by cross correlating the
stellar template acquired with SARG to a few suitable reference stars
observed with the same set-up and with available high-accuracy absolute
radial velocity from \cite{2002ApJS..141..503N}. It results in $-3.15\pm0.20$ km/s.

\section{Stellar parameters}
\label{s:star}

\subsection{Spectroscopic analysis}

We have used  the TNG/SARG template spectrum of  \hd to provide an
independent  assessment of its  atmospheric parameters  (\ensuremath{T_{\rm eff}}, \ensuremath{\log{g}},
and [Fe/H]) with respect to the values reported by~F07.

Our  methodology follows  a standard  procedure whose  details  can be
found     in     several     works     of     the     recent     past~
\citep[e.g.,][]{1996AJ....111..424G,2001AJ....121..432G,2004A&A...415.1153S}.
We briefly summarize
it here. We initially selected  a set of relatively weak \ion{Fe}{I} and
\ion{Fe}{II}  lines~ \citep[see, e.g.,][  and references  therein, for
details on  the line list]{2004ApJ...616L.167S}, and  measured
equivalent widths (EWs) using the automated software ARES
\footnote{\tt   http://www.astro.up.pt/$^\sim$sousasag/ares},  made
available to the community by \cite{2007A&A...469..783S}.
The EWs  measured with ARES  are then fed  to the 2002 version  of the
MOOG spectral synthesis code \citep{1973PhDT.......180S}
\footnote{\tt  http://verdi.as.utexas.edu/moog.html}, together with
a  grid  of  Kurucz  ATLAS plane-parallel  stellar  model  atmospheres
\citep{1993KurCD..13.....K}.

The  atmospheric  parameters  of   \hd  are  then  derived  under  the
assumption  of  local  thermodynamic  equilibrium,  using  a  standard
technique of  Fe ionization balance~\citep[see,  e.g.,][and references
therein]{2004A&A...415.1153S,2004ApJ...616L.167S}.
We obtained  \ensuremath{T_{\rm eff}} $=  6100  \pm 75$~K,
\ensuremath{\log{g}} $= 4.1  \pm 0.1$, and [Fe/H]$= +0.14 \pm  0.08$, the formal errors
on  \ensuremath{T_{\rm eff}}\,  and  \ensuremath{\log{g}}\,  having  been derived  using  the  procedure
described   in~\cite{1997A&A...328..261N}   and~\cite{1998A&A...339L..29G},
while   the
nominal uncertainty for [Fe/H]  corresponds to the scatter obtained from
the \ion{Fe}{I} lines rather than the formal error of the mean.

We also quantified the sensitivity of our iron abundance determination
to   variations  of   $\pm1\sigma$   with  respect   to  the   nominal
$T_\mathrm{eff}$, and $\log g$ values, and found changes in [Fe/H] of at
most 0.05 dex, below the adopted uncertainty of 0.08 dex.

\begin{figure}
\resizebox{\hsize}{!}{
\includegraphics{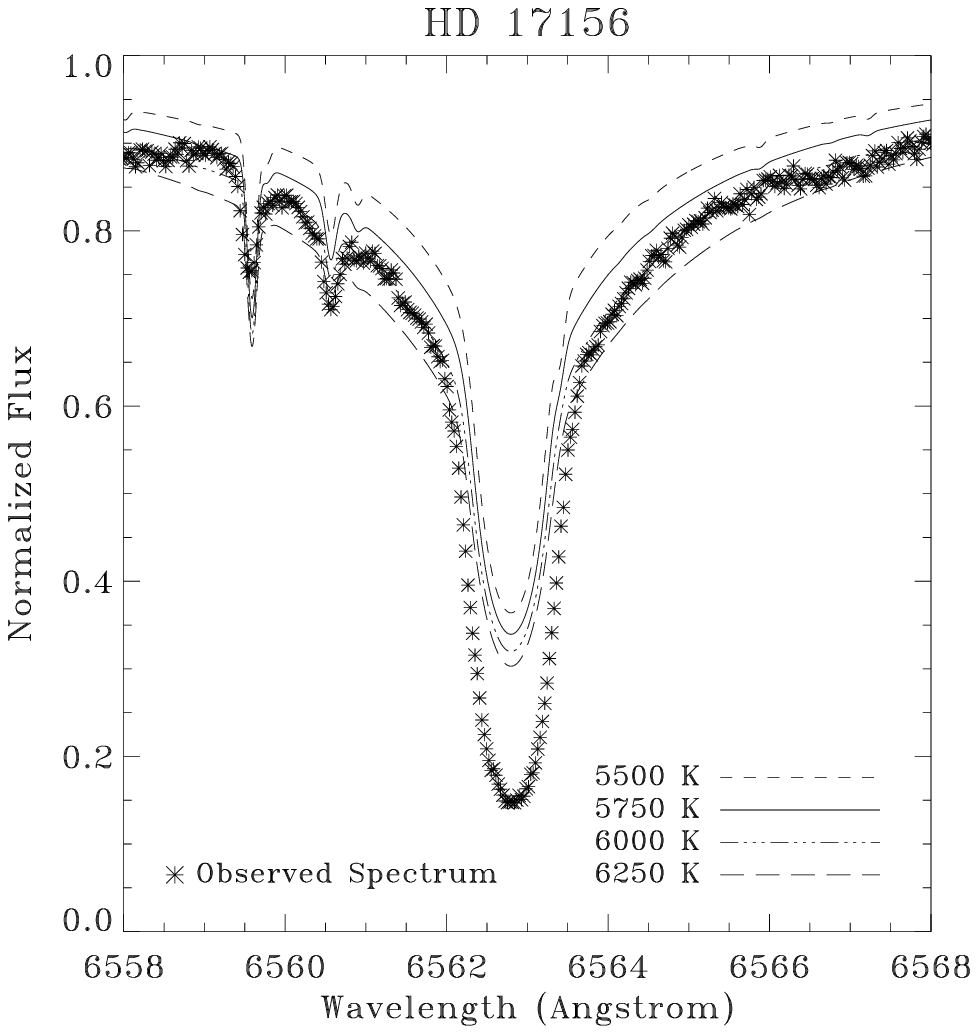}}
\caption{The portion of the spectrum of \hd around H$_\alpha$ and
synthetic profiles for four different \ensuremath{T_{\rm eff}}. A \ensuremath{T_{\rm eff}} close to 6000~K
is suggested by the fit of the wings of the line. }
\label{halpha}
\end{figure}

With  the  aim  of  further   testing  the  accuracy  of  the  \ensuremath{T_{\rm eff}}\,
determination above, we have  carried out a few additional consistency
checks. For example, in  Figure~\ref{halpha} we show the comparison of
the  observed  H$_\alpha$  line  profile  in  an  archival  Keck/HIRES
spectrum against four  synthetic profiles for solar-metallicity dwarfs
([Fe/H]  = 0.0,  $\log  g =  4.5$)  from the  Kurucz  database. As  is
well-known, the H$_\alpha$ line is very sensitive to changes in \ensuremath{T_{\rm eff}},
while relatively insensitive to changes in   \ensuremath{\log{g}}\,   and
[Fe/H]\,~\citep[see, e.g.,][and references therein]{2006A&A...458..997S,2007ApJ...664.1190S},
thus this exercise certainly  helps to test the accuracy of
the  spectroscopic  \ensuremath{T_{\rm eff}}\,  derived   above.  The  results  shown  in
Figure~\ref{halpha}, in which a  10~\AA\ region centered on H$_\alpha$
is  displayed together  with  four calculated  profiles for  different
\ensuremath{T_{\rm eff}}\,  values,  indicate rather good  agreement  with the  estimate
reported in Table~\ref{t:pstar}.

\subsection{Age}
\label{s:age}

Based on isochrone fitting, F07  reported an
age estimate for \hd of 5.7$^{+1.3}_{-1.9}$ Gyr,
suggesting an old, slightly evolved F8/G0 primary.
We performed an independent isochrone fitting using the set of isochrones of
\cite{2000A&AS..141..371G} and the software PARAM described in
\cite{2006A&A...458..609D}\footnote{\tt http://stev.oapd.inaf.it/param}.
The input values for PARAM are the parallax, the visual magnitude,
[Fe/H] and \ensuremath{T_{\rm eff}}.
We used the parallax from \cite{2007A&A...474..653V} and for the
metallicity and effective temperature we ran the code
twice, once with the values from F07 and once with
our estimate.
The results of PARAM for the stellar age are 2.4 $\pm 1$ and 2.8
$\pm 1$ Gyr, for the F07 and our parameters, respectively.
These are only marginally compatible with the previous age
estimate by F07.
The stellar mass and radius are instead fully compatible (see below).

Other indirect age indicators confirm an age of a few Gyr.
The low level of Ca II H\&K chromospheric activity suggests an age
of about 6 Gyr \citep{2007ApJ...669.1336F}.
The lack of X-ray emission from ROSAT \citep{2000IAUC.7432R...1V}
yields an upper limit of $ \log L_{X} < 28.7$.
This in turn implies an age older than 1.6 Gyr, using the
age-X ray emission calibration by \cite{2008ApJ...687.1264M}.

To obtain an additional age estimate and to investigate possible
chemical peculiarities of \hd with respect to other planet hosts with
similar physical properties, we have measured its lithium (Li)
abundance.

\begin{figure}
\resizebox{\hsize}{!}{
\includegraphics{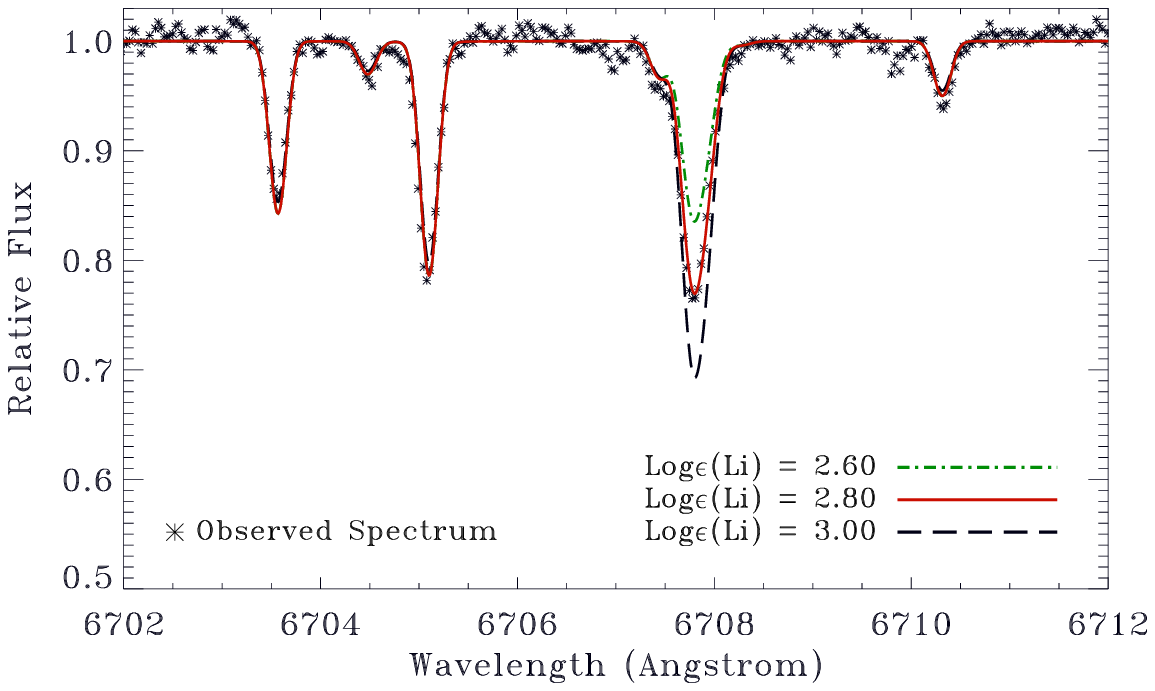}}
\caption{
Portion of \hd spectrum centered on Lithium 6707.8 \AA\ line.
Overplotted the results of the spectral synthesis for three different
lithium abundance.}
\label{f:lithium}
\end{figure}

Figure~\ref{f:lithium} shows a spectral synthesis of a 10~\AA\ region
centered on the Li ~$\lambda = $ 6707.8 \AA ~line in an archival Keck/HIRES
spectrum of \hd, and using the atmospheric parameters derived from the
Fe-line analysis and the line list of \cite{2002MNRAS.335.1005R}.
In the figure, the observed spectrum is compared to three synthetic
spectra, each differing
only in the assumed Li abundance. We find a best-fit value of
$\log\epsilon{\rm (Li)} \approx 2.80$ for \hd. We then infer a rather
old age for the star of $t > 2$ Gyr, based on the average Li abundance
curves as a function of effective temperature for clusters of different
ages reported by~\cite{2005A&A...442..615S}.

The measured Li abundance for \hd does not appear peculiar when compared
to that of sub-samples of nearby planet hosts with similar
\ensuremath{T_{\rm eff}}\,~\citep{2004A&A...414..601I,2008MNRAS.386..928G}.
To further investigate these
issues we will present in a future paper a more detailed study of the
elemental abundances in \hd.

\subsection{Stellar mass and radius}
\label{s:param}

F07 provided mass and radius  estimate from isochrone
interpolations (mass 1.2 $\pm 0.1$ M$_\odot$, radius 1.47$^{+0.13}_{-0.17}$~R$_\odot$).
Our isochrone fit (see Sect.~\ref{s:age}) gives
1.21 $\pm 0.03$ and 1.19 $\pm 0.03$ M$_\odot$ when assuming \ensuremath{T_{\rm eff}} and [Fe/H] from
the F07 and our analysis respectively, and
same radius 1.35 $\pm 0.08 $ R$_\odot$

The comparison of the results of the two fits show that they are
compatible within error bars, but the best value are slightly different. The
origin of these differences is not univocal, and to understand this problematic
independent measurements of the star mass and radii are needed.

\subsubsection{Monte Carlo experiment. Method}

For this purpose we adopted an approach
based  on   a  Monte  Carlo  experiment. In  each
realization we  generated a set  of observed stellar  properties.
Using using calibrations from the literature we obtained the corresponding
values  for  mass  and  radius.  From the  resulting  distributions  we
obtained the  most probable  values and relative  errors for the mass and
radius.

In more detail, we created 10$^6$ different synthetic systems where we
generated Gaussian distributions for the parallax, the
{\em V} and {\em K} magnitudes, and \ensuremath{T_{\rm eff}}, using as standard deviations their
respectively error bars.
The input  parameters and  relative their standard deviations  are reported  in
Tab.~\ref{t:pstar}.
In the same way, we generated values
for the bolometric correction and the bolometric magnitude of the Sun.
For  each synthetic set  obtained
this way  we calculated the absolute magnitudes,  luminosity, radius,
mass, density and gravity.

The  {\em V}   magnitude  was  obtained   from  Simbad, {\em K}
magnitude  from 2MASS  after  conversion to the
Bessel-Brett system \citep{2001AJ....121.2851C}, and the
\ensuremath{T_{\rm eff}}  from  our spectroscopic  determination.   For  the parallax  we
adopted the recently revised Hipparcos  value \citep{2007A&A...474..653V},
which indicates that  \hd lies at 75 pc from the  Sun, 3 pc closer
than  the  previous estimate. The bolometric correction was set to B.C.$=-0.03$ \citep{2000A&AS..141..371G}
while for the solar bolometric magnitude we used the value M$_{bol,\odot}= 4.77\pm$0.02.

The stellar  radius  was obtained  using  the  \ensuremath{T_{\rm eff}}-K mag calibration  of
\cite{2004A&A...426..297K},
 and  using the  Stefan-Boltzmann law. The  stellar mass
was  calculated from various  mass-luminosity relations  (MLR) namely:
{\it i)} the classical MLR $L  \propto M^{4.5}$; {\it ii)} the MLR of
\cite{2007MNRAS.382.1073M} using absolute magnitude and ;{\it iii)} stellar luminosity;
{\it iv)} the  MLR of \cite{2004ASPC..318..159H}.  Density and gravity
were estimated directly from the radius and mass assuming a spherically symmetric star.

\subsubsection{Monte Carlo experiment. Results}
\begin{figure}
\centering
\resizebox{\hsize}{!}{
\includegraphics[angle=-90]{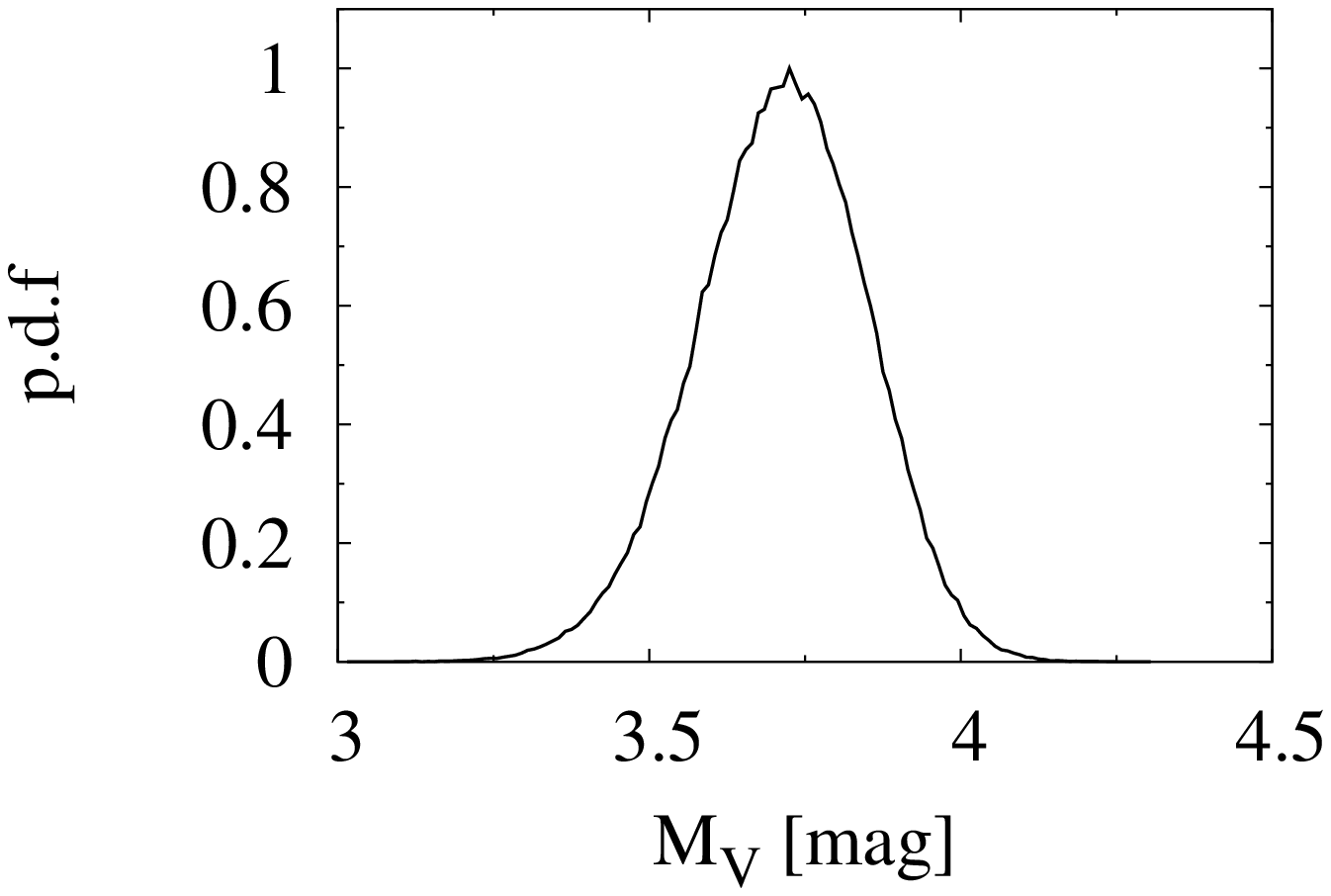}
\includegraphics[angle=-90]{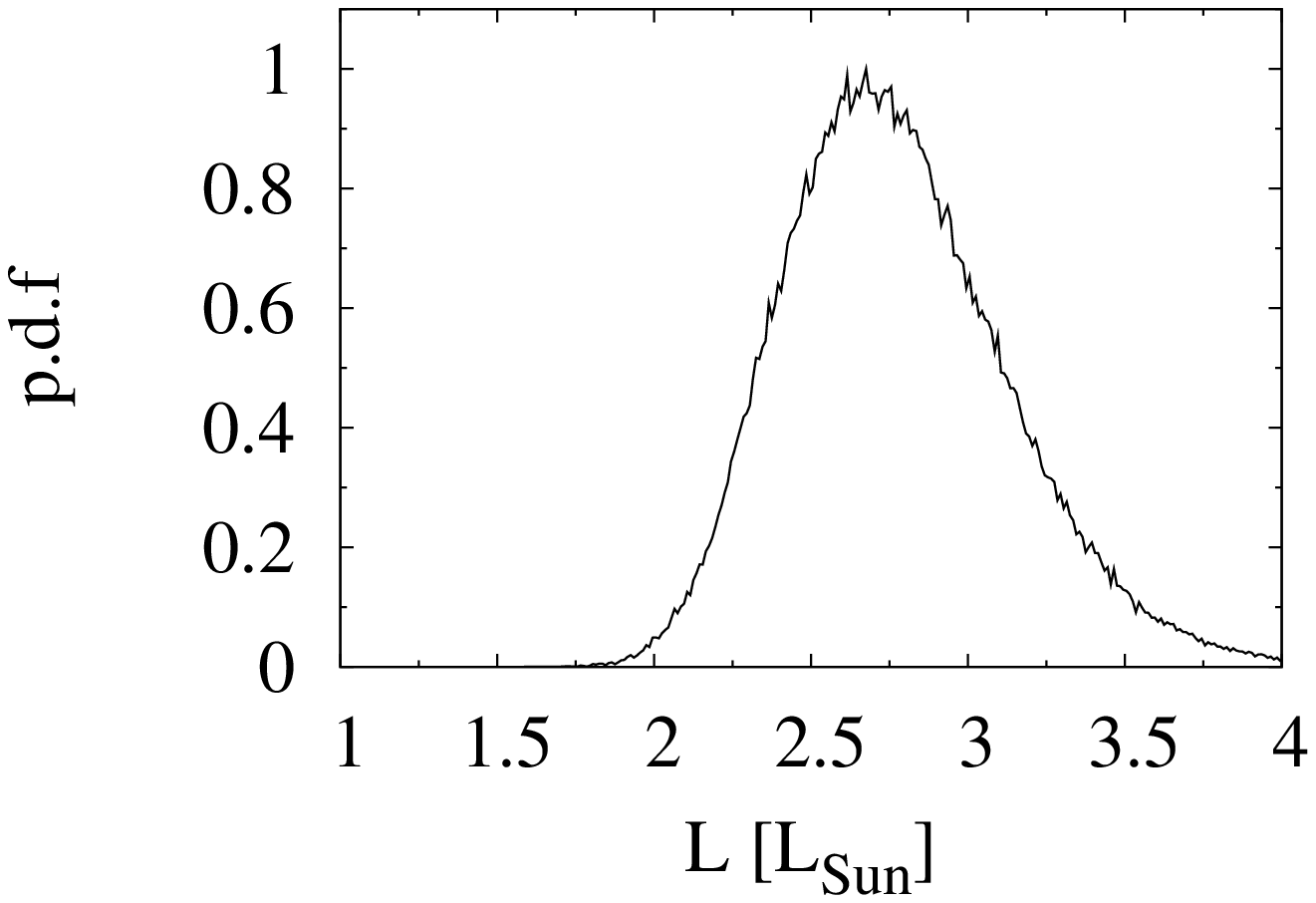}
}
\resizebox{\hsize}{!}{
\includegraphics[angle=-90]{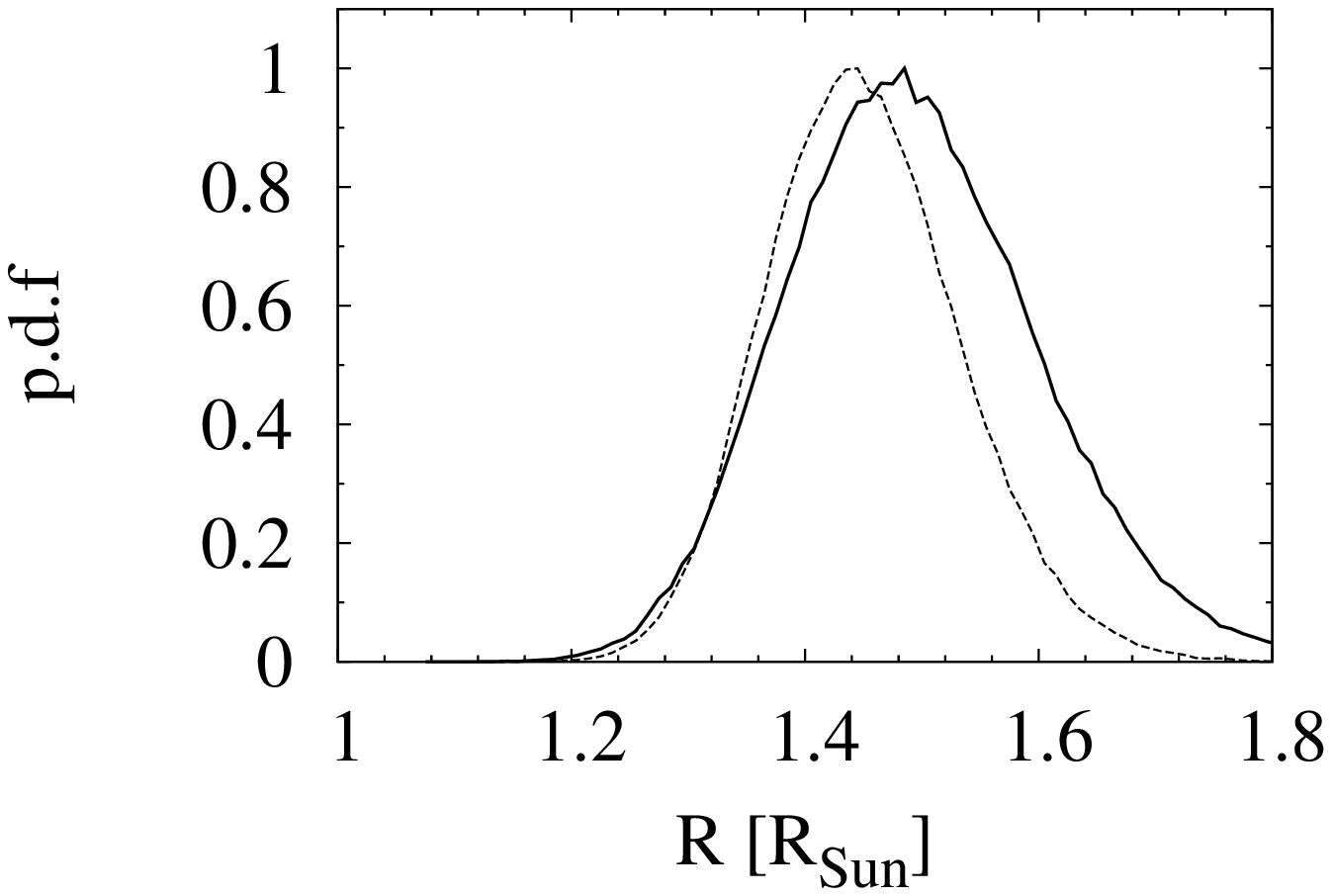}
\includegraphics[angle=-90]{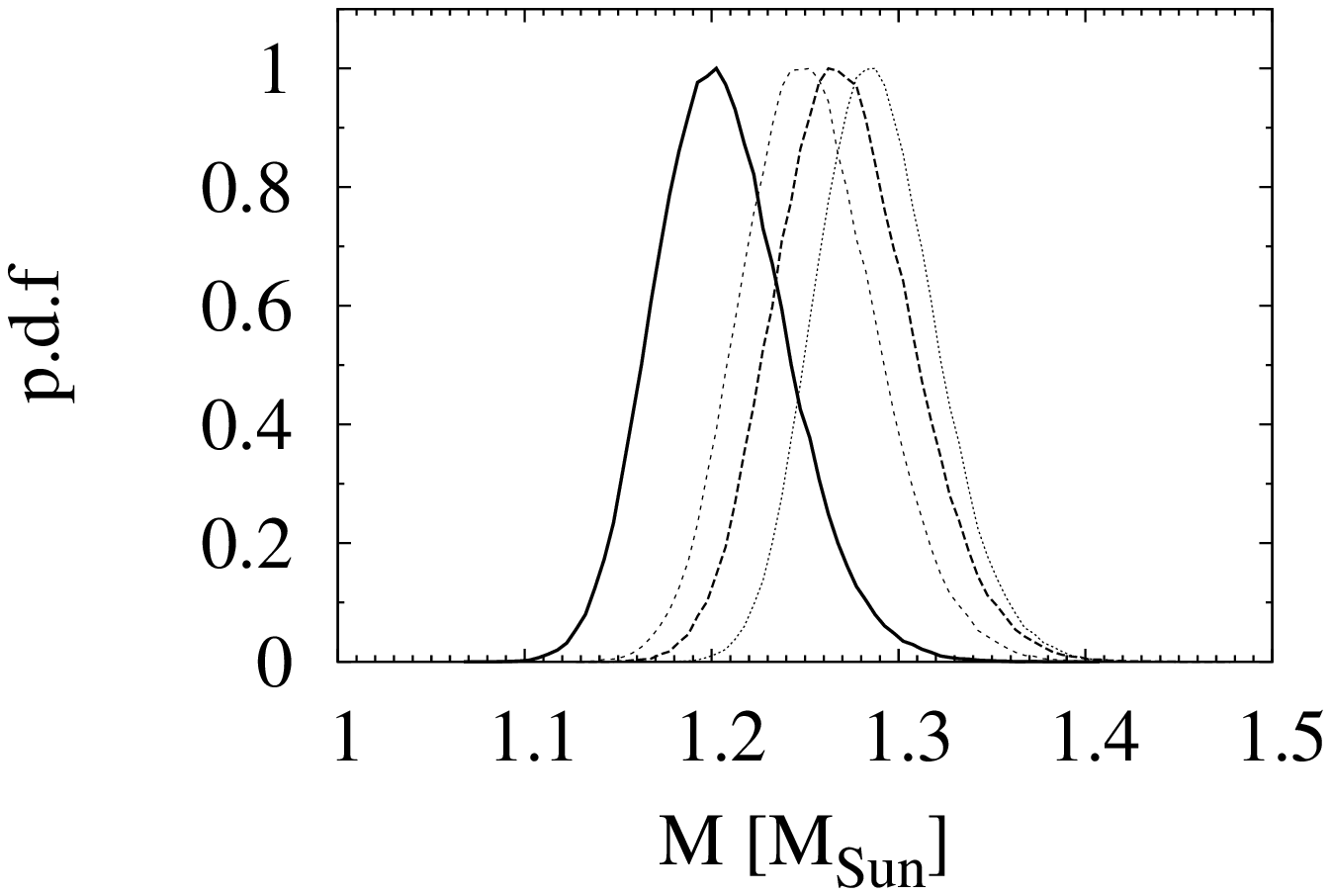}
}
\resizebox{\hsize}{!}{
\includegraphics[angle=-90]{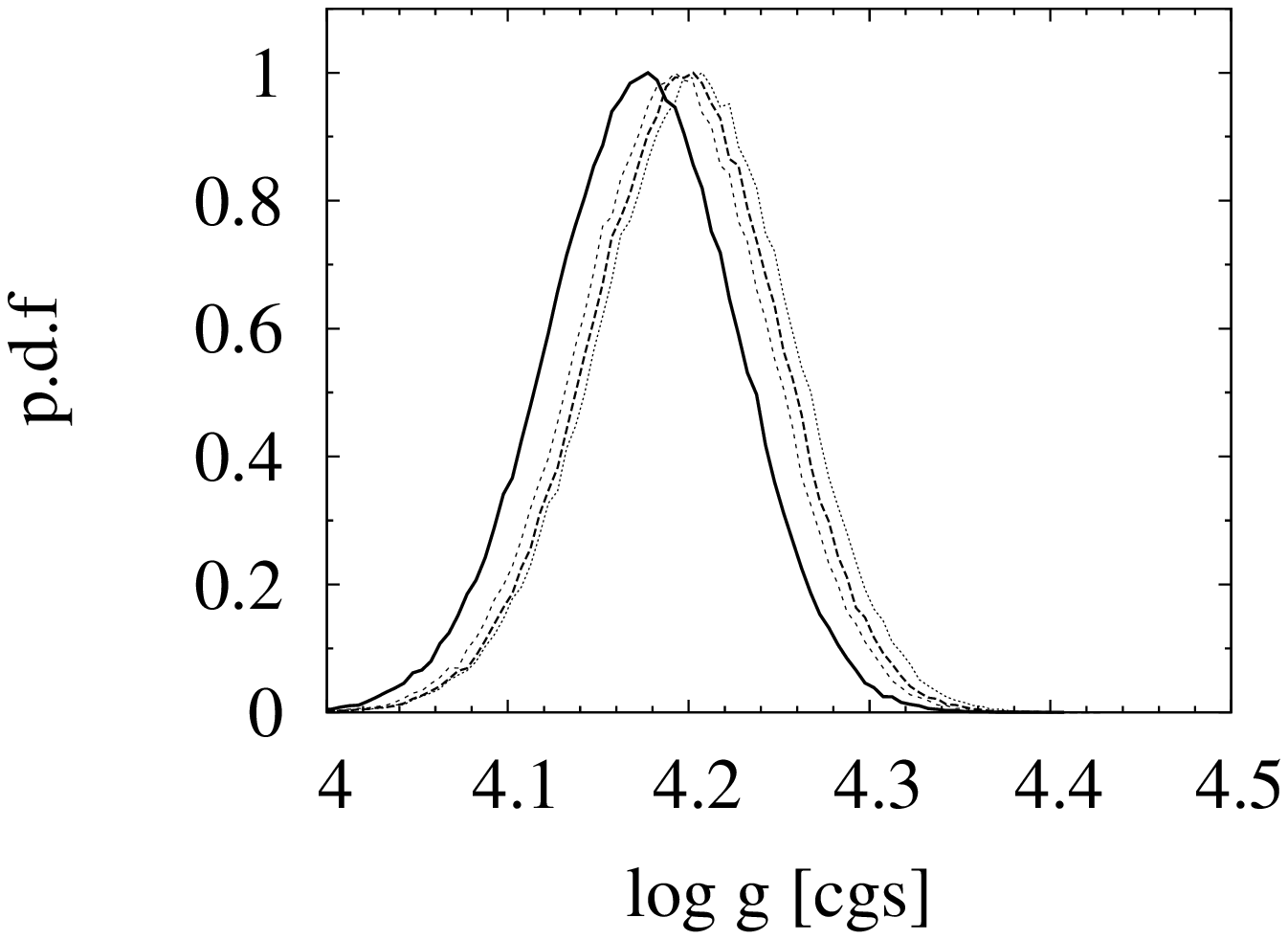}
\includegraphics[angle=-90]{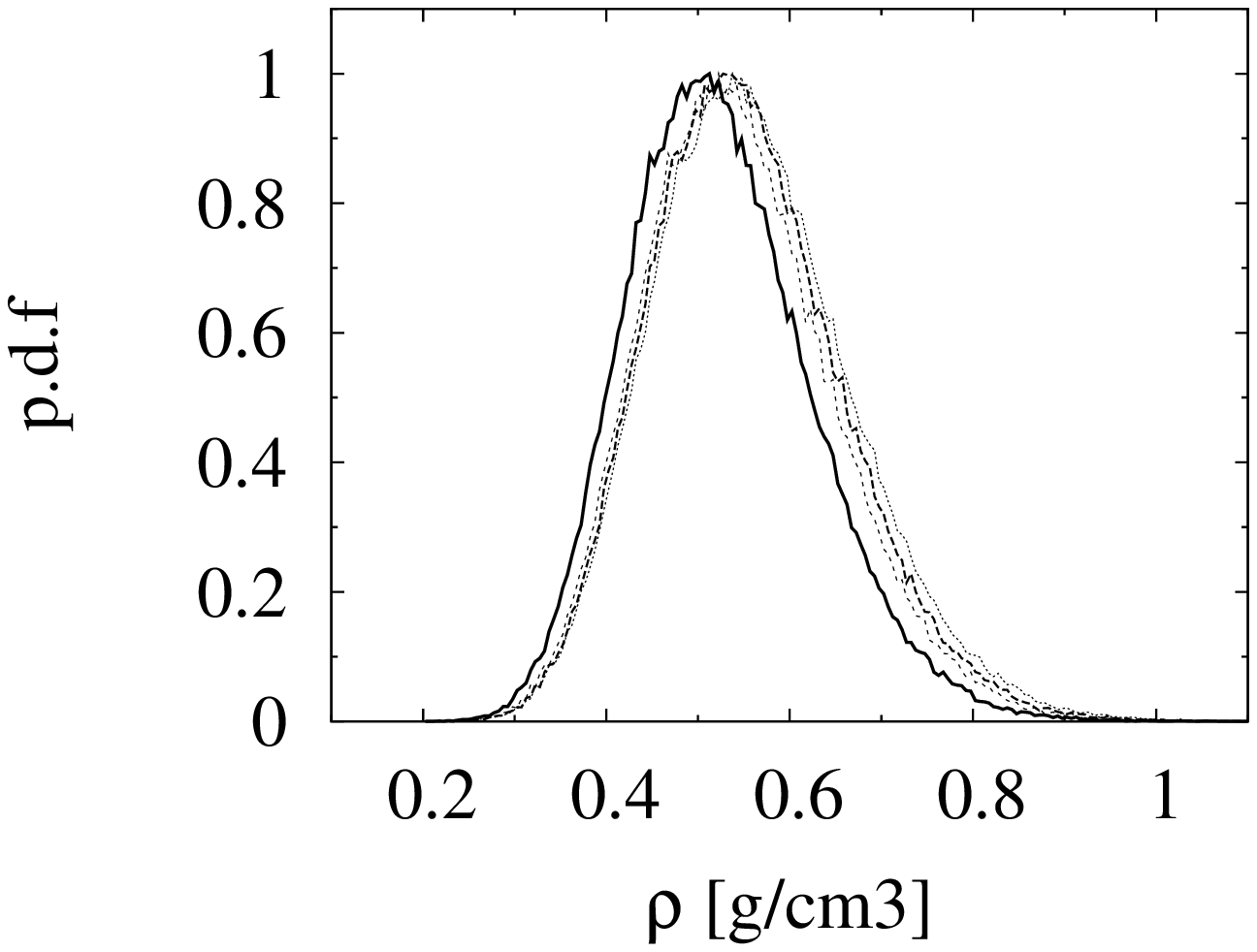}
}
\caption[]{Probability density function of the stellar parameters obtained with the Monte Carlo experiment.
{\it Upper left:}  Absolute magnitude
{\it Upper right:} Luminosity
{\it Middle   left:}  Radius: thick line results using Stefan-Boltzmann law, thin line results using the calibration of Kervella(\ensuremath{T_{\rm eff}},M$_{\rm K}$)
{\it Middle   right:} Mass, from left to right results using: Malkov MLR(M$_{\rm V}$), $M \propto L^{4.5}$, Malkov MLR(L), and Henry MLR(M$_{\rm V}$)
{\it Lower left:}  \ensuremath{\log{g}}
{\it Lower right:} Density
}
\label{f:distri}
\end{figure}

The  distribution  of  absolute  magnitudes  in  V  band  peaks  at  3.7
(Fig.~\ref{f:distri}, upper left), while  the luminosity is peaked at 2.5
L$_\odot$ (Fig.~\ref{f:distri}, upper right). These values are slightly
different from the  ones obtained by \citeauthor{2007ApJ...669.1336F}, and  the source of these
differences is ascribed only to the difference in the adopted parallax.

The   resulting   distributions   for   the   radius   are   shown   in
Fig.~\ref{f:distri}  (middle left).   The two  relations  used provide
similar results  (R$_S\sim 1.4$R$_\odot$) for  the best value  and also the
shape of the  distributions is very similar. \citeauthor{2007ApJ...669.1336F}  suggest a slightly
larger radius. Also in this case, the difference originates from the
change in the adopted parallax.

In Fig.~\ref{f:distri} (middle  right) we present  the mass distributions
obtained with the MLRs.  Using the relation
of \cite{2004ASPC..318..159H} we obtain the  highest mass
(M $\sim 1.28  $M$_\odot$). However, we note that this relation was originally
derived  from  parameters  of close  binary  stars.  \cite{2003A&A...402.1055M}
demonstrated that this  kind of relations do not  well describe single
stars. In order to avoid this problem  we adopted the MLRs  of
\cite{2007MNRAS.382.1073M} obtained on detached main-sequence double-lined
eclipsing binaries. These
relations are also valid for slightly evolved stars  like \hd (almost
all the stars used for deriving these relations are also slightly evolved.
O.~Malkov, private communication). We obtain best  values for the  mass between
1.2  and 1.24  M$_\odot$ ,  the first  obtained using  MLR($M_V$)  and the
second using MLR($L$). The classical MLR ($M \propto L^{4.5}$) provides
M $\sim $1.22  M$_\odot$. These  results are  consistent with  the values
estimated by F07 (1.2 $\pm$ 0.1 M$_\odot$).

Finally, the  resulting distributions for  the gravity and  the density
are  portrayed in  the lower  panels of  Fig.~\ref{f:distri}.   The mean
values are  \ensuremath{\log{g}} = 4.22  and $\rho$ =  0.58 g\,/cm$^3$.

The results  of this  experiment show a fairly good  agreement  with the
estimate of R$_S$ based on isochrone fitting (both our and the one by
F07). We conclude that the two independent
approaches based on isochrone fitting and the use of scaling relations provide
consistent results.

In the following we do not adopt a  value for the radius,
because  we want to determine  independently its  value from the light-curve
fits. Moreover,  we fix  the value of the mass to
the value of the weighted mean of our mass estimation (not using the Henry MLR results)
M =  1.24$\pm$0.03 M$_\odot$.
We summarize in Table \ref{t:pstar} all the data relative to
this Monte Carlo experiment.

\begin{table}
\centering
\caption{
{\it Upper panel} Input parameters for the Monte Carlo experiment and best values.
{\it Lower panel} Kinematical properties and galactic orbit parameters.
}
\label{t:pstar}
\begin{tabular}{lll}
\hline
\hline
\multicolumn{3}{c}{Monte Carlo experiment} \\
\multicolumn{3}{c}{\it Input parameters} \\
\hline
parallax                &     13.33$\pm$ 0.72    &  mas                 \\
mag V                   &    8.172 $\pm$ 0.031   &  mag                 \\
mag K                   &    6.807 $\pm$ 0.024   &  mag                 \\
\ensuremath{T_{\rm eff}}                   &     6100 $\pm$  75     &  K                   \\
B.C.                    &    -0.03 $\pm$ 0.02    &  mag                 \\
\multicolumn{3}{c}{\it Output parameters} \\
M$_{\rm bol,\odot}$     &     3.69 $\pm$ 0.12    &  mag                 \\
M$_{\rm V}$             &     3.73 $\pm$ 0.12    &  mag                 \\
L                       &     2.68 $\pm 0.28$    & L$_\odot$                \\
R (Stefan-Boltzmann)    &     1.49 $\pm 0.09$    & R$_\odot$                \\
R (Kervella)            &     1.45 $\pm 0.07$    & R$_\odot$                \\
M (Malkov,M$_{\rm V}$)  &     1.21 $\pm 0.04$    & M$_\odot$                \\
M (Malkov,L)            &     1.25 $\pm 0.04$    & M$_\odot$                \\
M ($M \propto L^{4.5}$) &     1.22 $\pm 0.04$    & M$_\odot$                \\
M (Henry,M$_{\rm V}$)   &     1.29 $\pm 0.03$    & M$_\odot$                \\
\ensuremath{\log{g}}    (mean)         &     4.21 $\pm 0.05$    & cgs                  \\
$\rho$   (mean)         &     0.57 $\pm 0.10$    & g\,/cm$^3$                \\
\hline
\multicolumn{3}{c}{Kinematical properties} \\
RV                  &  -3.15   $\pm$   0.2  &  \ensuremath{\rm km\,/s}               \\
$\mu_\alpha$        &  91.14   $\pm$   0.49 &  mas/yr             \\
$\mu_\delta$        & -33.14   $\pm$   0.56 &  mas/yr             \\
U                   &   0.6    $\pm$   0.2  &  \ensuremath{\rm km\,/s}               \\
V                   &  26.1    $\pm$   2.0  &  \ensuremath{\rm km\,/s}               \\
W                   & -22.8    $\pm$   1.5  &  \ensuremath{\rm km\,/s}               \\
R$_{\rm min}$       &   8.0    $\pm$   0.3  &  kpc           \\
R$_{\rm max}$       &  10.9    $\pm$   0.3  &  kpc           \\
R$_{\rm med}$       &   9.5    $\pm$   0.4  &  kpc           \\
Z$_{\rm max}$       &   0.2    $\pm$   0.1  &  kpc           \\
e (galactic orbit)  &   0.15   $\pm$   0.05 &                \\
\hline
\end{tabular}
\end{table}

\subsection{Rotational velocity}
\label{s:vsini}

F07 derived v$\sin i=2.6\pm 0.5$ km/s for \hd.
Our template spectrum is suitable for an independent measurement of
this quantity.
We discuss here three different methods adopted for
the measurements of v$\sin i$, based on our template spectra.

\begin{figure}
\resizebox{\hsize}{!}{
\includegraphics{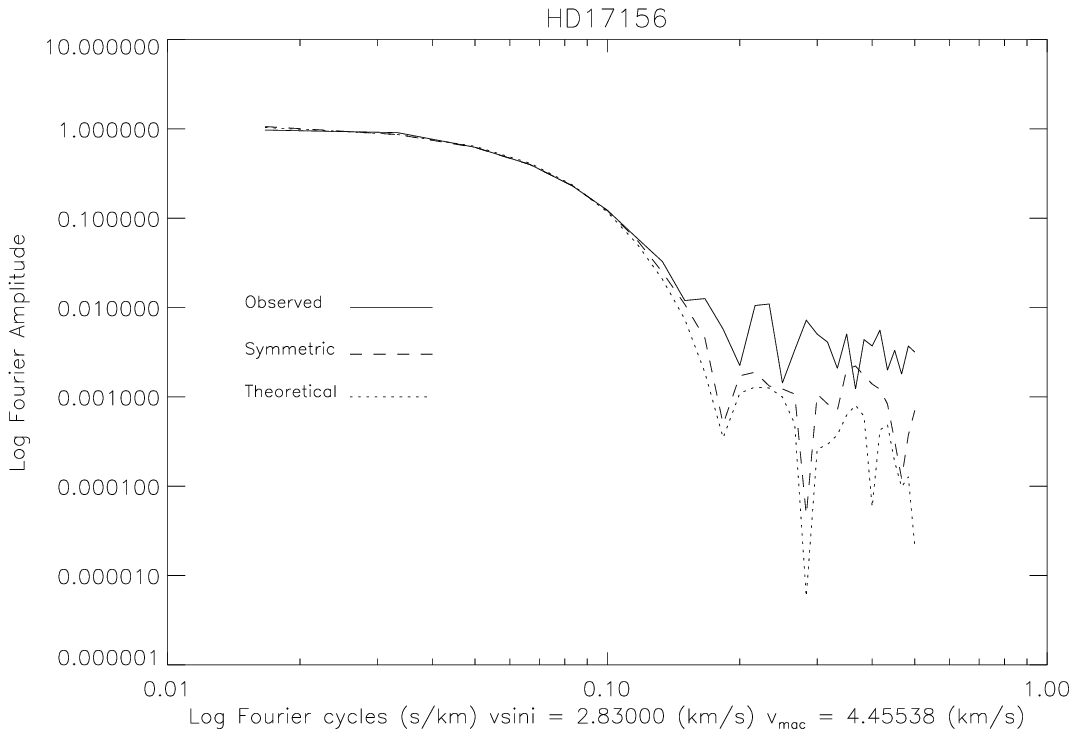}}
\caption{
Estimation of v$\sin i$ through the FFT. Three profiles were analyzed: the
observed profile (solid line), the symmetric profile (dashed line)
mirroring one half of the profile, and a theoretical profile (dotted
line). The FFT corresponds to a case considering $T_{eff}=6079$ K and Eq.\,1
given by \citep{2005ApJS..159..141V}.
}
\label{f:vsini_aldo}
\end{figure}

With the first method we derive v$\sin i$ by the Fast Fourier Transform
analysis of the star's absorption profile
(see Fig.\ref{f:vsini_aldo}).
To determine the $v \sin i$, the observed profile of a stellar absorption line is made
symmetric by mirroring one of its halves, with the purpose to reduce
the noise of the FFT. A new profile is calculated by the convolution
of a macroturbulence profile (Gaussian) and a rotational one, to compare
the FFTs of the symmetric and the calculated (model) profiles.
The v$ \sin i$  value of the rotation profile, is set as variable parameter
until the first minimum of the FFT from the calculated profile coincides
with the minimum of the FFT from the symmetric one.
The value of v$\sin i$ for \hd was determined considering possible values of
macroturbulent velocity ($v_{mac}$) from B-V and \ensuremath{T_{\rm eff}} \citep{2005ApJS..159..141V},
and we obtained a v$\sin i$ ranging from 1.8 to 2.8 \ensuremath{\rm km\,/s}.

The second method  that we used consists in obtaining the rotational velocity by means of a
suitable calibration of the FWHM of the cross-correlation
function against the B-V color.  This relation was derived for all the stars
in the SARG planet search survey, and it was calibrated into v$\sin i$ using
stars with known rotational velocity from the literature.
Using the B-V from the Tycho catalog converted to the Johnson system
(B-V=0.632), the resulting v$\sin i$ $= 3.2 \pm 1$ \ensuremath{\rm km\,/s} .

Finally, using MOOG we synthesized a number of isolated \ion{Fe}{I} lines in
the template spectrum. From these we measured  v$\sin i$ $= 3.0 \pm 0.5$ \ensuremath{\rm km\,/s}.

The values obtained with the three methods suggest a range of values for
v$\sin i$ compatible with the measurement of F07. The measurement of
v$\sin i$ from the analysis of Rossiter-Mclaughlin effect will be
presented in Section \ref{s:rm}.

\subsection{Galactic orbit}

The measurements of the absolute radial velocity given in Section
\ref{s:hrs}, together with the revised parallax and proper motion from
Hipparcos \cite{2007A&A...474..653V}, allow one to calculate the space velocity
with respect to the Local Standard of Rest and the galactic orbit of \hd. Space velocities are
calculated following the procedure delineated by \cite{1987AJ.....93..864J} and \cite{1989A&A...218..325M},
adopting the value of standard solar motion of \cite{1998MNRAS.298..387D}
(with U positive toward the galactic anticenter). The calculations
yield $(U,V,W) = (+0.6,+26.1,-22.8)$ \ensuremath{\rm km\,/s}.
The galactic orbit of the
star is obtained integrating the equations of motion of a massless
particle in the potential described by \cite{1991RMxAA..22..255A}. The
equations of motion are solved using the RADAU integrator \citep{1985dcto.proc..185E}
assuming that the rectangular galactocentric coordinates of the
Sun are $(X_\odot,Y_\odot,Z_\odot)=(8.0,0.0,0.015)$ kpc, and that the
local circular velocity is 220 km s$^{-1}$. We compute 1\,000  orbits each
time varying randomly the initial coordinates and velocity of the star
within the error bars and integrating for a time of 4 Gyr ($\approx$ 15
full orbits around the Galactic Center).
The UVW spatial velocity and the mean values of the computed orbits are
reported in Table \ref{t:pstar}. The mean galactic radius provides
an estimate of the galactocentric distance of the star at the moment of
its formation. The value for \hd, $R_\mathrm{med} = 9.5$ kpc, implies
that the star spends most of its time in regions outside of the solar circle.

\section{Photometric observations}
\label{s:photom}

Photometric observations of the transit of 2007 December 3 were obtained
with various telescopes. On 2007 December 2 we used three medium-class
telescopes (Asiago 1.82 m and OHP 1.20 m) and a number
of small telescopes, including six 30-40 cm amateur-operated
telescope all located in continental Europe as well as
the Telast 0.3-m telescope in the Canary Islands. Weather conditions across
continental Europe were not optimal.

Observations at Asiago started under photometric conditions, transit
ingress was observed but observations ended at December 04 UT 02:30,
due to the presence of clouds, which prevented the observations of the third and
fourth contacts.
OHP observations were performed under
variable sky conditions due to intermittent clouds and veils, good observing
conditions were achieved only during the transit time window, with only a
small coverage of the Out Of Transit (OOT) flat part of the lightcurve
before the first contact. Observations with Telast were obtained under
normal sky conditions and were performed throughout the night.
Amateur observations were carried out by six observatories spread
over central and northern Italy. Observing conditions suffered from clouds and
veils as the others European sites involved in the campaign, and the full transit was
successfully observed by four telescopes, the remaining two observatories
obtaining data only for the ingress phase of the transit.

Three additional attempts to coordinate transit observations of \hdb
were carried out on 25 December 2007 and on September 25 and October 17 2008.
Unfortunately, due to bad weather conditions, there was only one useful observation
for each date.

All observations were obtained in {\em R} band except the two obtained by B.~Gary
that was acquired in white light; the characteristics of the
different telescopes are summarized in Table~\ref{t:telescopes}.

\subsection{Data reduction}
\label{s:lc_creation}

The Asiago, OHP, and Telast raw images were calibrated using flat field and
bias frames. The resulting images were analyzed with IDL routines to perform
standard aperture photometry. The center of the aperture was calculated using
a Gaussian fit, and the aperture radius was held fixed for each set
(in the range 15-20 pixels).
The sky background contribution was removed after an estimation
The brightest non-variable stars in the field
were measured in the same way, and a reference light
curve was constructed by adding the flux of these stars. The target flux
was divided by this reference to get the final normalized light-curve.

The images obtained by amateur telescopes were reduced  bias subtracted,
and dark flat-field corrected using commercial software.  Aperture
photometry of \hd was then performed using IRIS
\footnote{\tt http://www.astrosurf.com/buil/iris} and adopting a
fixed aperture equal to 2 times the stellar FWHM.
The sum of the flux of  the brightest  stars  in the field  was  used
as reference  for building the normalized light-curve.

The final  light-curve for each  telescope was corrected for
differential airmass and residual  systematic effects dividing them by
a linear  function of time to the region outside the transit.
The  photometric error on each  point of a
lightcurve was  calculated as the rms  over an interval  of 30 minutes
(the timescale  of the ingress/egress  phase). The typical rms  of the
OOT lightcurves  are   reported  in Table~\ref{t:telescopes}, while
the complete photometric dataset will be available in electronic format at CDS.

The whole dataset consists of $\sim$ 7\,000 photometric points.
We used these data  to perform a global analysis of  the planetary
transit.

\begin{figure}
\centering
\resizebox{\hsize}{!}{\includegraphics{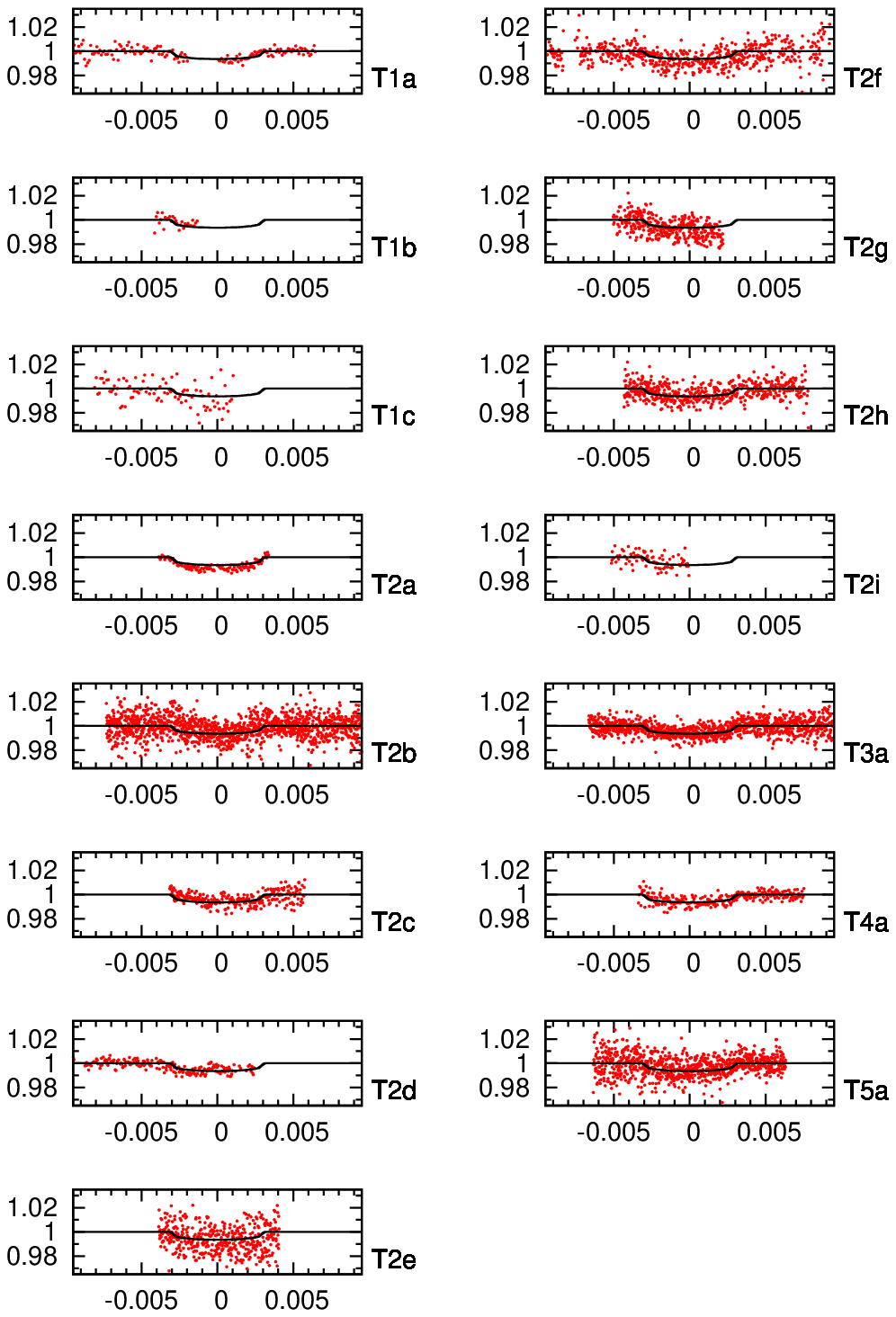}}
\caption{
Mosaic of the differential light curves obtained during transits of \hdb.
obtained with several telescopes.
In each box the horizontal axis is the photometric phase and the vertical axis is
the relative flux, along with the best fit model.
From left to right column and from top to bottom:
T1a=Almenara, T1b=Gasparri and T1c=Lopresti datasets from the \cite{2007A&A...476L..13B}.
T2a=OHP,
T2b=Telast,
T2c=Castellani,
T2d=Asiago,
T2e=Lopresti,
T2f=Marchini,
T2g=Nicolini,
T2h=Papini,
T2i=Vallerani.
T2a to T2i light-curves were collected on 2007 December 03.
T3a=Gary 2007 December 25,
T4a=Gregorio 2008 September 25,
T5a=Gary 2008 October 2008,
}
\label{f:ph}
\end{figure}

\begin{figure*}
\centering
\resizebox{\hsize}{!}{\includegraphics{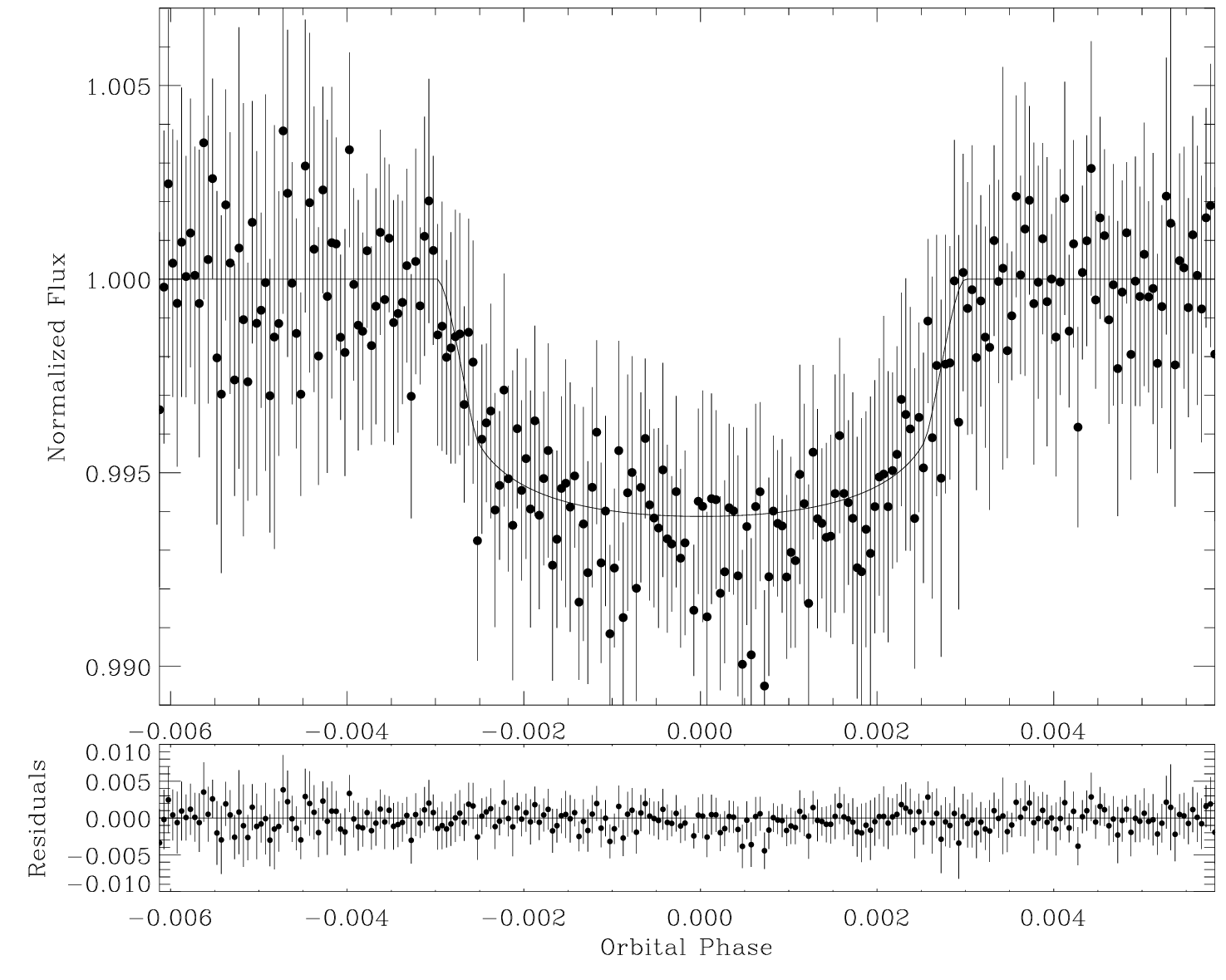}}
\caption{
Combined light curve of all photometric data folded with the
orbital period, along with the best fitting model.
The time interval between each point is 90 s.
}
\label{f:ph_phased}
\end{figure*}

For the light curve fitting we used all the lightcurves that we have
collected without performing data binning.
In Fig.\ref{f:ph} are portrayed the light curves used in this study,
folded with the best orbital period from the fit.
For displaying  purposes the combined light-curve of the 15 lightcurves
is shown in Fig.~\ref{f:ph_phased}. This combined light curve was obtained using a bin width of 90~s,
the OOT has an rms of 0.0016.

\begin{table}
\caption{Summary of photometric observations of \hd performed on 2007/08.
The last column report the midtransit times for each date obtained from the fit in Sec~\ref{s:lc_analysis}.
}
\label{t:telescopes}
\resizebox{\hsize}{!}{
\centering
\begin{tabular}{l|l|l|l|llllllll}
\hline
\hline
Site/Observer  &  diameter   &  rms OOT  &  date                 &  T$_c$                         \\
               &  m          &  mmag     &                       &  BJD                           \\
\hline
Telast         &    0.30     &      4.2  &    2007 September 10  &  2454353.61300     $\pm$ 0.0200\\
Gasparri       &    0.20     &      5.3  &                       &                                \\
Lopresti       &    0.18     &      7.0  &                       &                                \\
\hline
Asiago         &    1.82     &      2.8  &    2007 December 03   &  2454438.47450     $\pm$ 0.0005\\
OHP            &    1.2      &      2.0  &                       &  \\
Telast         &    0.3      &     11.2  &                       &  \\
Obs. Cavezzo   &    0.4      &      6.1  &                       &  \\
Lopresti       &    0.18     &     10.5  &                       &  \\
Obs. Univ.Siena&    0.3      &      8.4  &                       &  \\
Obs. Mt. Baldo &    0.4      &      7.3  &                       &  \\
Papini         &    0.3      &      6.9  &                       &  \\
Vallerani      &    0.25     &      5.9  &                       &  \\
\hline
Gary           &    0.3      &      5.6  &    2007 December  25  &  2454459.69087     $\pm$ 0.0032\\
\hline
Gregorio       &    0.3      &      3.8  &    2008 September 25  &  2454735.51351     $\pm$ 0.0027\\
\hline
Gary           &    0.3      &      8.9  &    2008 October   17  &  2454756.73134     $\pm$ 0.0024\\
\hline
\end{tabular}
}
\end{table}

\section{System parameters}
\label{s:system}

We performed the analysis of the \hd system in three steps.
First, using the radial velocities presented in Table 1 along with
other published RV values : F07 (2 datasets: Keck+Subaru),
\cite{2008PASJ...60L...1N}, \cite{2008ApJ...683L..59C} (2 datasets: HET + HJST),
we have derived a new spectroscopic orbital solution for \hd.
A full Keplerian orbit of five parameters:
the radial velocity semi-amplitude K$_{RV}$,
the time of periastron passage T$_P$,
the orbital period $P$, the orbital eccentricity $e$,
and the argument of periastron $\omega$ was adjusted to the data.
Second, we have carried out a fit to all the ligh-curve that we have obtained and
also to the \citet{2007A&A...476L..13B} datasets using the $e$, and the $\omega$
obtained from the orbital solution.
In this scheme, the adjustable parameters
are the ratio of the radii $k=R_p/R_s$ their relative sum $(R_s+R_p)/a$,
the orbital inclination $i$, the midtransit time $T_c$, and the orbital period $P$.
The values derived from the light-curve analysis were then used to determine,
through the analysis of the Rossiter-McLaughlin, new values of v$\sin i$ and of the angle $\lambda$ between the equatorial plane of the
star and the orbital plane of the planet.

The modeling of the transit lightcurve and Rossiter-McLaughlin
effect was carried out using the analytical formulae provided by
\cite{2006A&A...450.1231G} and \cite{2006ApJ...650..408G}.
The mathematical basis for the description of the two effects is
the same, i.e. the \cite{1977Ap&SS..50..225K} theory of eclipsing binary stars.
This fact warranties an internal consistent description
of the observed data.

\subsection{Orbital radial velocity analysis}
\label{s:rv_analysis}

In order to derive the stellar spectroscopic orbit using the
combined set of radial velocities mentioned above we used only the OOT
measurements in all datasets.
Observations in the night of the transit are valuable to this goal
because of the steep RV slope (about 23 m/s/hr). We use a downhill
simplex algorithm to perform the RV fit to the six datasets, including the zero
point shifts between the datasets as free parameters.
A stellar jitter
of 3 m/s was added in quadrature to the observational errors F07.

\begin{figure}
\centering
\resizebox{\hsize}{!}{\includegraphics[angle=-90]{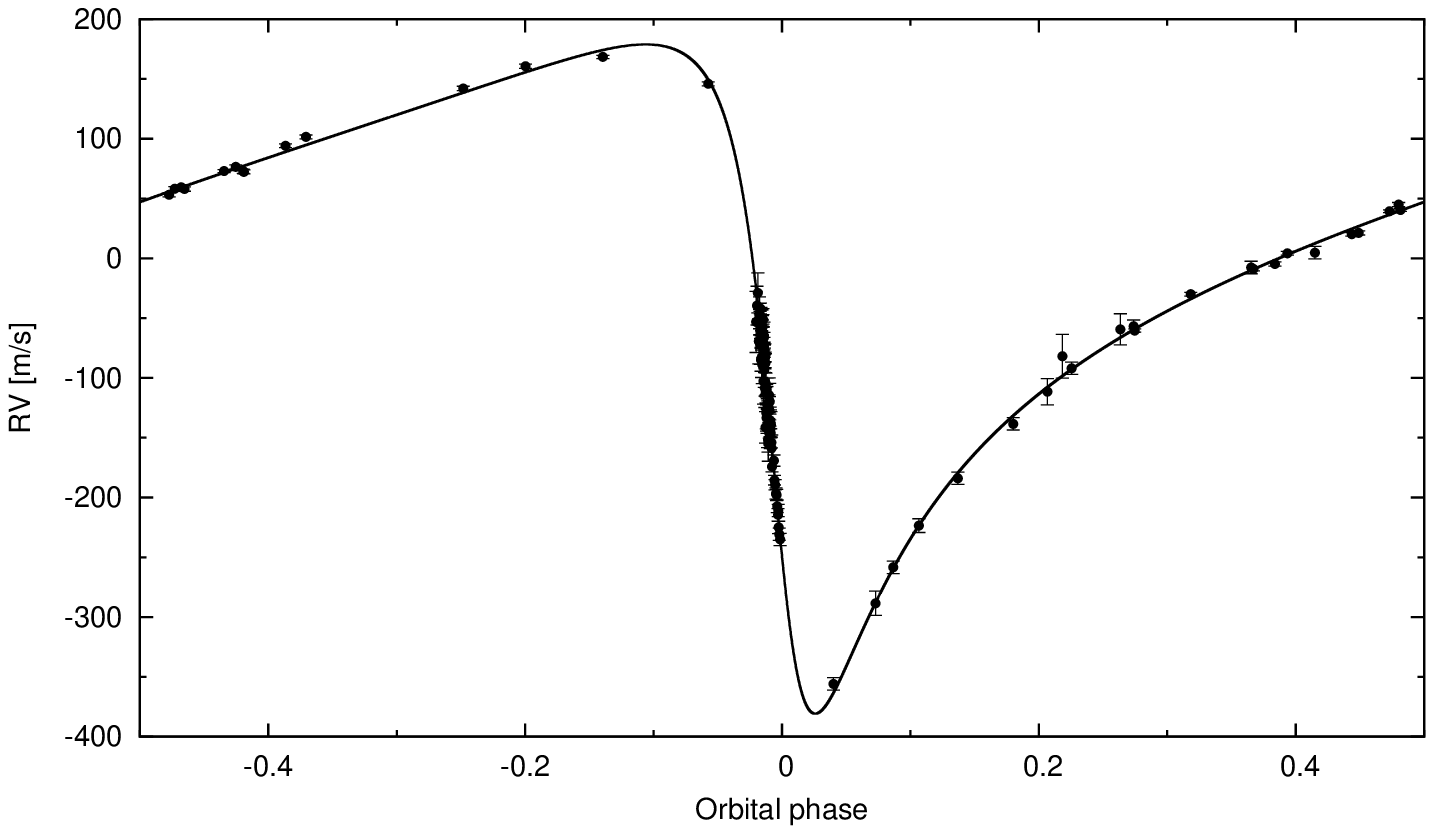}}
\resizebox{\hsize}{!}{\includegraphics[angle=-90]{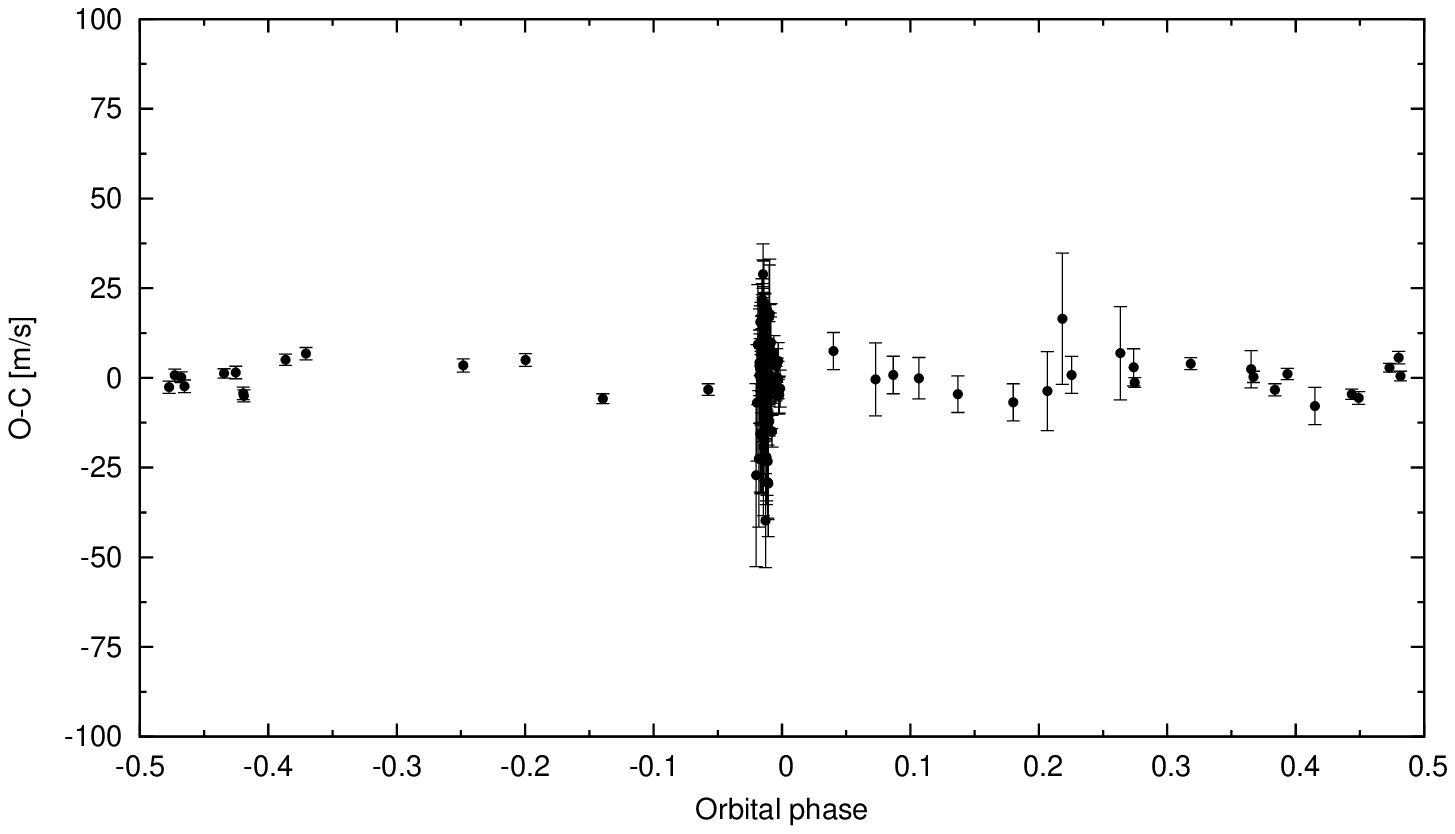}}
\caption{
Upper panel: radial velocities of \hd phased to the best-fit orbital solution.
Bottom panel: residuals from the orbital solution.
}
\label{f:rv}
\end{figure}

The best-fit solution has a value of reduced $\chi^2 = 1.13$,
and the results are in close agreement with the discovery paper
F07 and its subsequent analysis
\citep{2008ApJ...681..636I}. Uncertainties in the
best fit parameters were obtained exploring the $\chi^2$ grid with an
adequate resolution. The orbital solution and relative parameter
uncertainties are presented in Table~\ref{t:plan_param}. In Figure~\ref{f:rv} we show
the phased radial velocity curve with the best-fit model. Using the
value of primary mass provided in Section \ref{s:param} and its uncertainty,
the resulting minimum mass for the planet is $m\sin i = 3.21 \pm 0.08$ M$_{\rm J}$, and
the semi-major axis is $a = 0.1614 \pm 0.0010$ AU.

\begin{table}
\caption{Parameters of the \hd system.
}
\label{t:plan_param}
\centering
\begin{tabular}{lll}
\hline
\hline
\multicolumn{3}{c}{{\it Orbital parameters}} \\
P               &       21.21663    $\pm$ 0.00045 &   day         \\
a               &        0.1614     $\pm$ 0.0022  &   AU          \\
e               &        0.682      $\pm$ 0.0044  &               \\
$\omega$        &        121.9      $\pm$ 0.23    &   deg         \\
$\lambda$       &        4.8        $\pm$ 5.6     &   deg         \\
K$_{RV}$        &        279.8      $\pm$ 0.06    &   m/s         \\
T$_P$           &  2454757.00787    $\pm$ 0.00298 &  BJD         \\
phase ingress   &       -0.003144   $\pm$ 0.00034 &              \\
phase egress    &        0.003160   $\pm$ 0.00034 &               \\
transit duration&        3.21       $\pm$ 0.08    &   hour        \\
\hline
\multicolumn{3}{c}{{\it Star parameters}} \\
M$_S$           &        1.24       $\pm$  0.03   &   M$_{\odot}$ \\
R$_S$           &        1.44       $\pm$  0.08   &  R$_\odot$        \\
L               &        2.58       $\pm$  0.36   &  L$_\odot$        \\
$\log\,g_S$     &        4.22       $\pm$  0.05   &  cgs       \\
$\rho_S$        &        0.59       $\pm$  0.06   &  g/cm$^{3}$       \\
v$\sin i$&   1.5        $\pm$  0.7    &  km/s              \\
\hline
\multicolumn{3}{c}{{\it Planet parameters}} \\
M$_P$           &        3.22       $\pm$  0.08   &   M$_{\rm Jup}$ \\
R$_P$           &        1.02       $\pm$  0.08   &  R$_{\rm J}$        \\
$\log\,g_P$     &        3.89       $\pm$  0.06   &  cgs       \\
$\rho_P$        &        3.78       $\pm$  0.06   &  g/cm$^{3}$       \\
\hline
\end{tabular}
\end{table}

\subsection{Photometric analysis}
\label{s:lc_analysis}

We used the description of  \cite{2006A&A...450.1231G}
to analyze the light-curves obtained in Sect.~\ref{s:lc_creation}.
In our model we allowed to vary the ratio of the radii, the phase of first
contact, the time of transit center and
the orbital inclination.
We fixed the limb darkening coefficients to the values
corresponding to a star with similar temperature
and metallicity to \hd from \cite{2000A&A...363.1081C} tables.
For the R band the adopted limb darkening coefficients were :
$u_+=u_a+u_b=0.6323$ and $u_-=u_a-u_b=-0.0655$.
The eccentricity, and the longitude of periastron were held fixed to
the best fit values obtained from the RV analysis (Sec. \ref{s:rv_analysis}).
Errors were estimated using the bootstrap scheme described in \citep{2008A&A...487L...5A}

The results of of the analysis of the 15 datasets are collected in Table~\ref{t:plan_param}.
Using the third Kepler's law we obtain for the stellar and planetary radii
$ R_S = 1.44 \pm 0.08 $ R$_\odot$ and  $ R_P = 1.02 \pm 0.08 $ R$_{\rm J}$.

\begin{figure}
\centering
\resizebox{\hsize}{!}{\includegraphics[angle=-90]{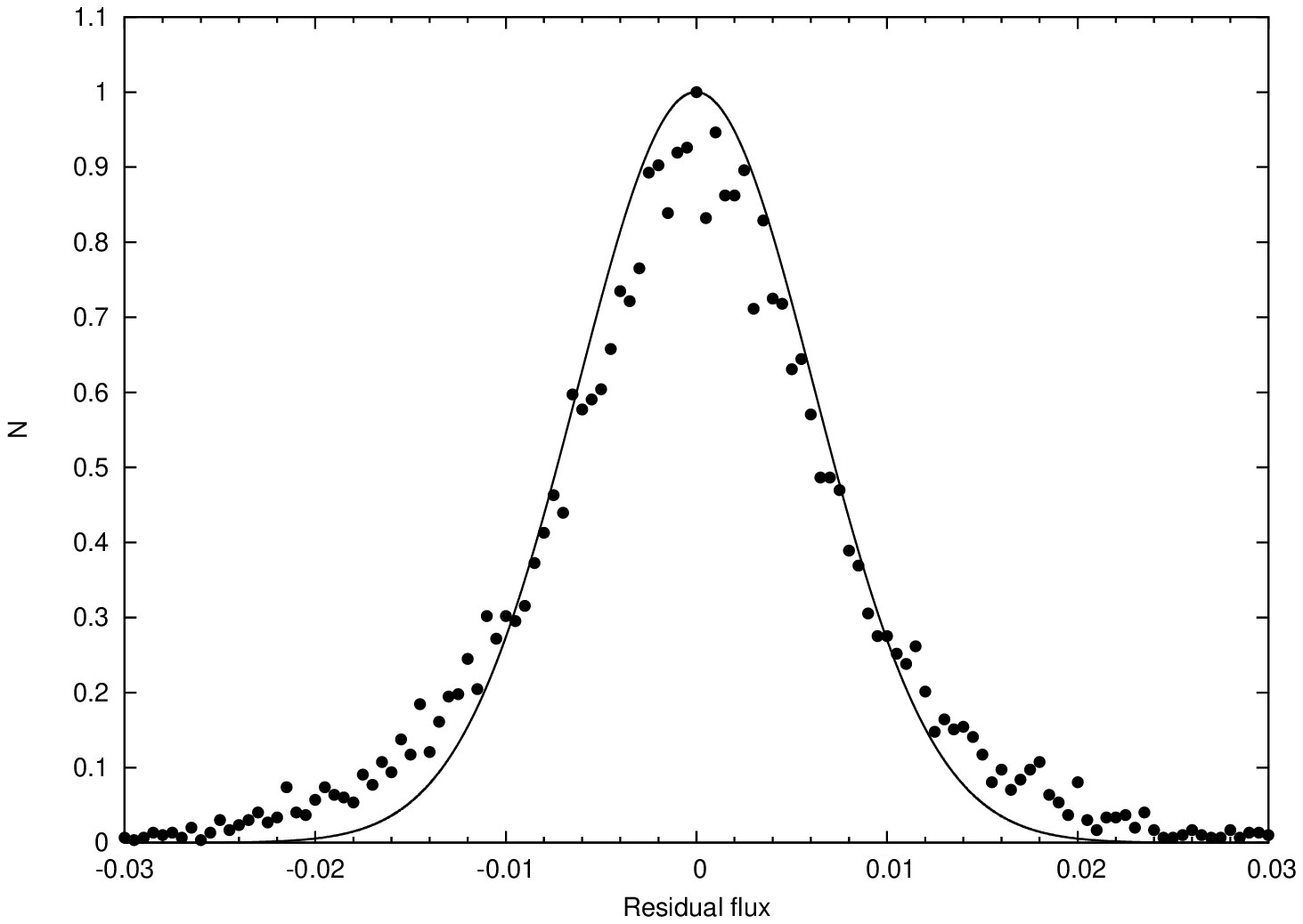}}
\caption{
Histogram of the lightcurve residuals. Overplotted the best Gaussian fit.
}
\label{f:ph_res}
\end{figure}

The histogram of the residuals of the light curves (Fig.~\ref{f:ph_res})
has a gaussian shape with standard deviation of 0.0062.

The    results   are    consistent   with    previous   determinations
\citep{2007A&A...476L..13B,2008ApJ...681..636I,2008PASJ...60L...1N,2008A&A...485..871G},
nevertheless  the values  of inclination  and  stellar  radius show  larger
deviations with respect to values presented by other authors.

In  their  analysis \cite{2008ApJ...681..636I} and \cite{2008PASJ...60L...1N} fixed the  stellar
radius to the value proposed by F07, while \cite{2008A&A...485..871G}
obtained  directly  the  radius   from  their  analysis.
The  origin  of the discrepancy on the stellar radius
might lie  in the different  values for the  inclination, because
its value controls the  transit duration and  the relative sum  of the radii.
For confirmation  we repeated the fit with  only the published
light-curve of  \cite{2008A&A...485..871G} and keeping the  limb-darkening coefficients
for the B band fixed to $u_+=u_a+u_b=0.7989$ and $u_-=u_a-u_b= 0.3019$.  The
results  are the  following:
$k =  0.0729  \pm 0.0031  $,
$\theta_1  = -0.00316 \pm  0.00023 $,
$i  = 87.9 \pm  0.1 $,
$T_c = 2454438.48372 \pm 0.00053 $ BJD,
R$_S = 1.44 \pm 0.07 $  R$_\odot$ and
$ R_P =  1.02 \pm 0.07 $ R$_{\rm J}$.
These  results are very close to the results of our previous fit.

We note that a transit model that does not take into account the non-zero eccentricity might
lead to erroneous results in the orbital inclination and thus also the stellar radius
(see, for instance, section 3.2 in \citealt{2008A&A...487L...5A}).

\subsection{Rossiter-McLaughlin effect analysis}
\label{s:rm}

\begin{figure}
\centering
\resizebox{\hsize}{!}{\includegraphics[angle=-90]{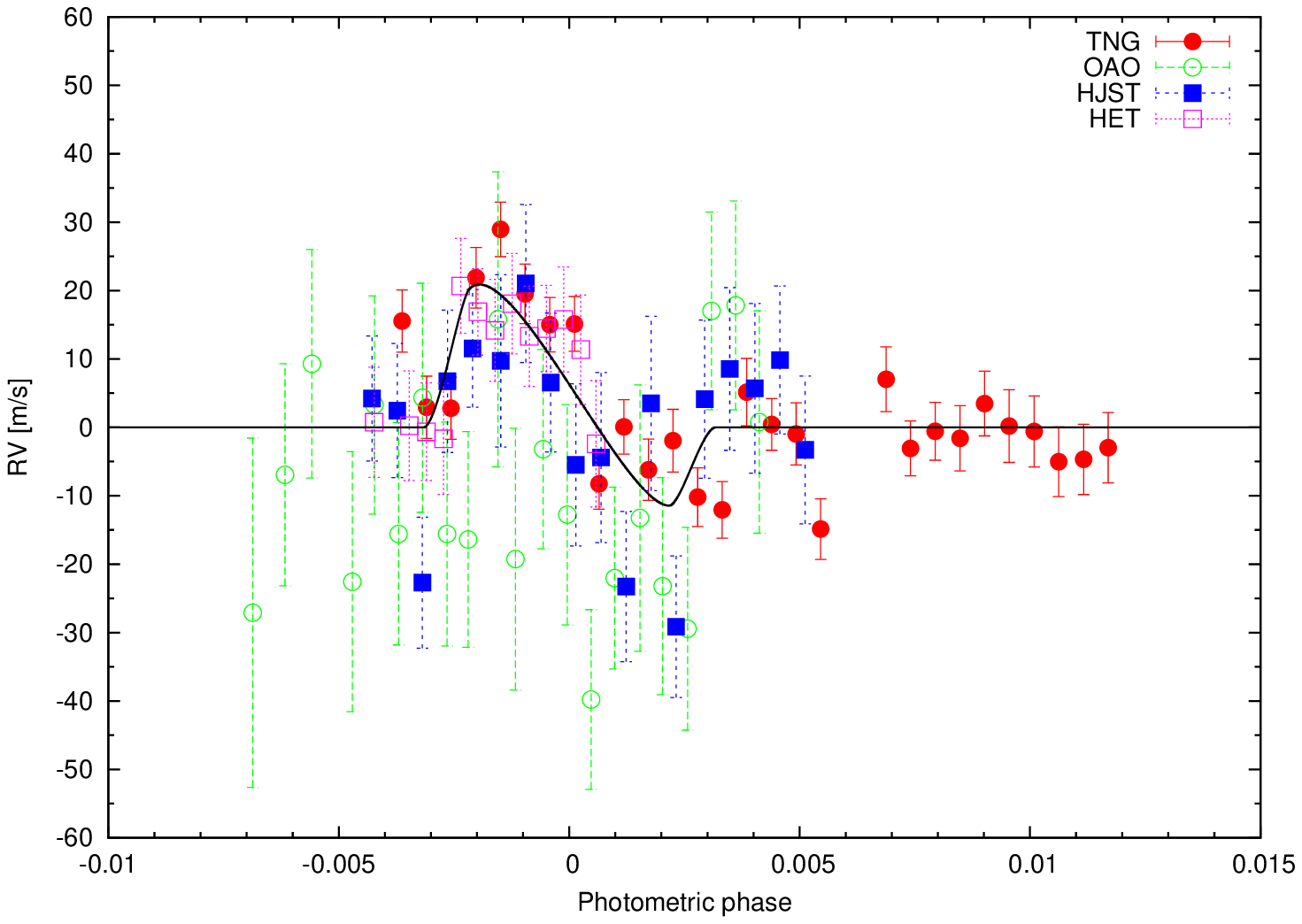}}
\resizebox{\hsize}{!}{\includegraphics[angle=-90]{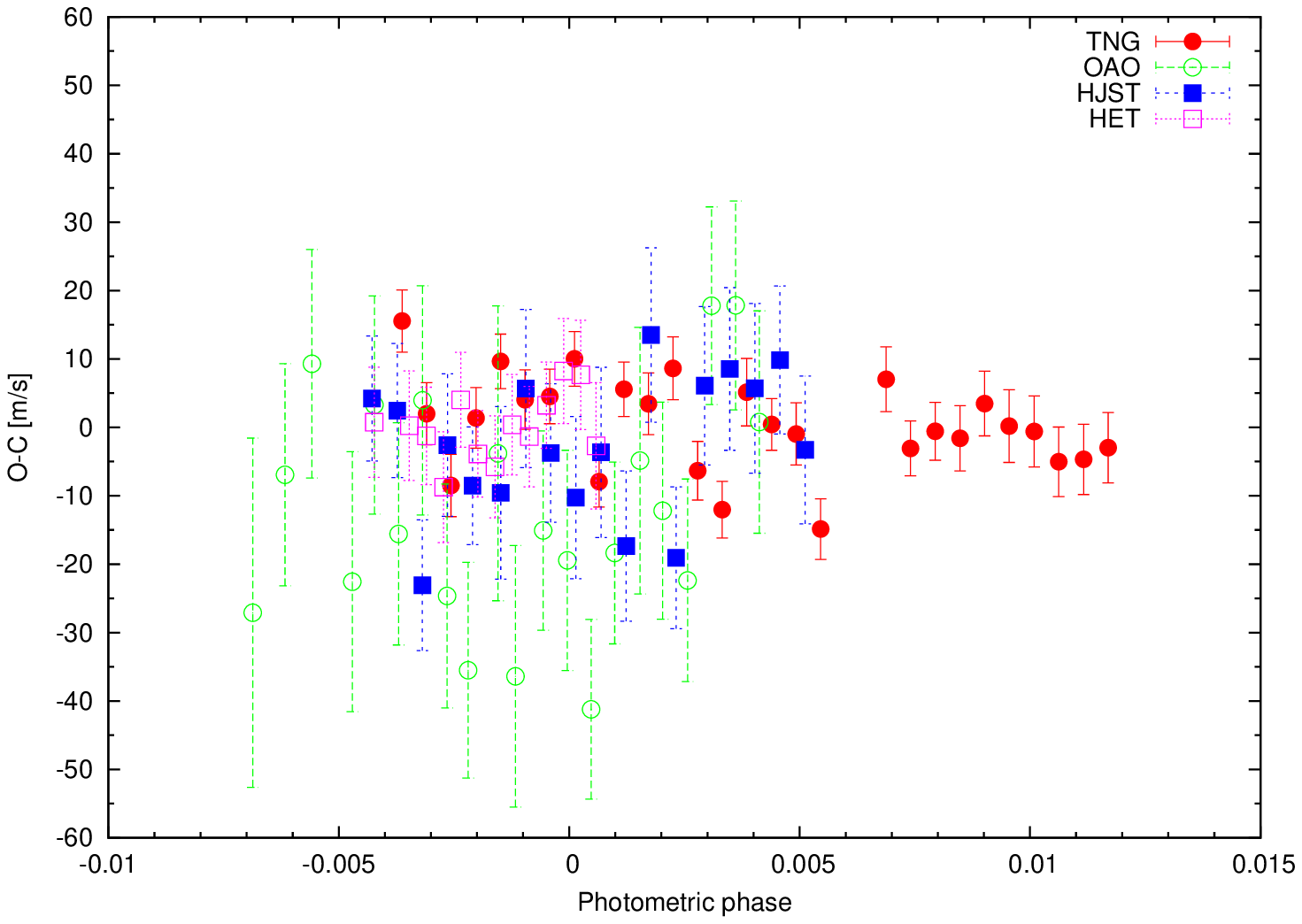}}
\caption{
Upper panel: residuals of radial velocities of \hd phased to the best-fit orbital solution
with the best fitted Rossiter-Mclaughlin effect overlayed.
Bottom panel: residuals from the Rossiter-Mclaughlin effect.
}
\label{f:rm}
\end{figure}

The analysis of the TNG RV data obtained during the transit was performed
using the formalism developed by \cite{2006ApJ...650..408G}.
We allowed to vary v$\sin i$ and $\lambda$ and we fixed
the values of $K$, $P$, $e$, $\omega$, $k$, $i$, $T_c$, $\theta_1$
to the best values obtained from RV and photometry analysis and reported
in Table \ref{t:rm}. Fig.\ref{f:rm} presents the best fitted model to the data.

The best fitted values to the RV orbital residuals of SARG are
v$\sin i$ $= 1.5 \pm 0.7$ \ensuremath{\rm km\,/s} and $\lambda = 4.8^\circ \pm 5.3^\circ $.
The value of v$\sin i$ agrees with the values determined
by F07 and by our analysis of the stellar spectra.
$\lambda$ is consistent with zero, indicating that the eclicptic plane
of the planet is closely aligned with the equatorial plane of the star.
This value of $\lambda$ does not confirm the claim of
\cite{2008PASJ...60L...1N}
for a large misalignment in this system, but rather agrees with
the relative alignment obtained by \cite{2008ApJ...683L..59C}.
Moreover, the two groups
find very different values for v$\sin i$. To study the nature of this discrepancy
we repeated the fit on their datasets independently, and the results of the fits
are summarized in Table \ref{t:rm}.
The results obtained using the HET dataset,
in spite of their good precision, do not provide strong constraints due to
their partial coverage of the transit and the fact that the zero point in the OOT
data cannot be estimated correctly. Instead, the fit to the HJST data are in excellent
agreement with the determination obtained with SARG.
Finally, the Narita et al. dataset provides a v$\sin i$ in good agreement with previous
determinations, and a value of $\lambda$ that formally points toward the
occurrence of some misalignment.
These results indicate also that our adopted description is consistent with the one
used by Narita et al. however, the intrinsically lower precision
of their RV data makes these results not significant.

\begin{table}
\caption{Results of the Rossiter-McLaughlin modeling of all datasets.
(OAO: Narita et al dataset).}
\label{t:rm}
\centering
\begin{tabular}{lrrrr}
\hline
\hline
telescope    &  date         &   v$\sin i$         &   $\lambda$        & $\chi^2$ \\
             &               &   \ensuremath{\rm km\,/s}           &   deg              & \\
\hline
OAO          &  17/11/2007   &   1.8 $\pm$ 1.5  &  -22.6 $\pm$ 20.3  & 0.7 \\
TNG          &   3/12/2007   &   1.5 $\pm$ 0.7  &    4.8 $\pm$  5.3  & 1.4 \\
HJST         &  25/12/2007   &   2.2 $\pm$ 1.0  &    0.8 $\pm$  6.4  & 0.9 \\
HET          &  25/12/2007   &   1.0 $\pm$ 5.0  &   30.1 $\pm$ 25.6  & 0.3  \\
\hline
TNG+HJST     &               &   1.6 $\pm$ 1.0  &    3.9 $\pm$  5.4  & 1.2 \\
\hline
\end{tabular}
\end{table}

\begin{figure}
\resizebox{\hsize}{!}{
\includegraphics{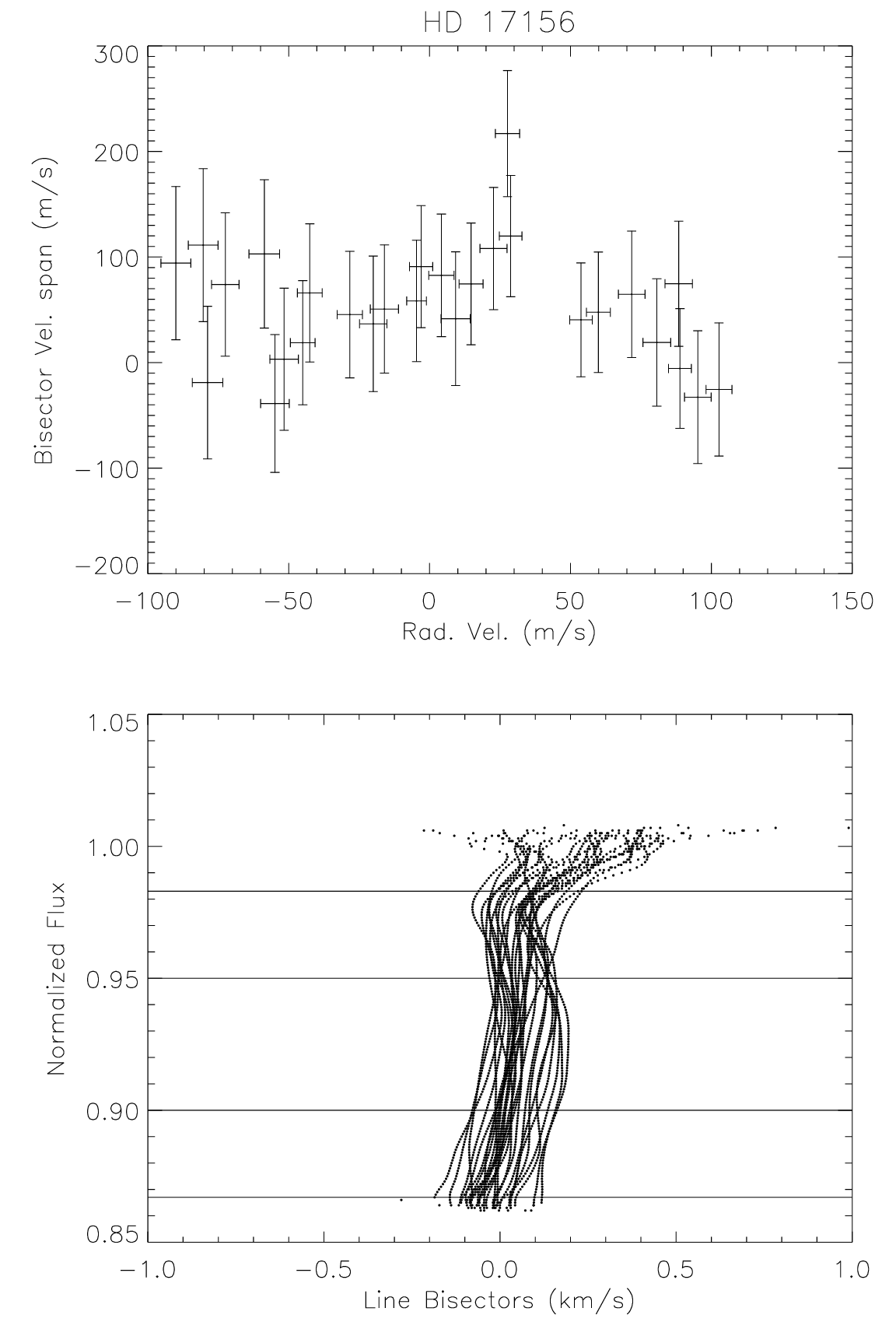}}
\caption{
Bisector velocity span (BVS) from line bisectors of HD~17156 spectra. Upper
panel displays BVS vs. RV. Lower panel shows the line bisectors from all
spectra, where the horizontal lines enclose the top and bottom zones
considered to compute the BVS.
}
\label{f:bvs}
\end{figure}

We also measured line bisectors and bisector velocity span using the
technique developed by \cite{2005A&A...442..775M} and looked
for changes in the line profiles caused by the planetary transit (see
\citealt{2008A&A...481..529L}). We do not detect significant variations
(Fig.~\ref{f:bvs}).
This is not unexpected considering the
typical signal to noise ratio of our spectra and the low amplitude of
the Rossiter signature.

\section{Clues on additional companions}
\label{s:binarity}

\hd was observed with AdOpt@TNG, the adaptive optics module of TNG
\citep{2006SPIE.6272E..77C}. The instrument feeds the HgCdTe Hawaii
1024x1024 detector of NICS, the near infrared camera and spectrograph of
TNG, providing a field of view of about 44 $\times$ 44 arcsec, with a
pixel scale of 0.0437 \arcsec/pixel. Plate scale and absolute detector
orientation were derived
in a companion program of
follow-up of binary systems with long term radial velocity trends from
the SARG planet search \citep{2007arXiv0705.3141D}.

Series of 15 sec images on \hd were acquired on 3, 18 and 23 October
2007 in Br$\gamma$ intermediate-band filter. Images were taken moving
the target in different positions on the detector, to allow sky
subtraction without the need for additional observations, and each night
at three different field orientations to make it easier to disentangle
true companions from image artifacts. The target itself was used as
reference star for the adaptive optics. Observing conditions were poor on
the night of October 3, and rather good on the nights of 18 and 23 October,
when we obtained a typical Strehl Ratio of about 0.3.

Data reduction was performed by first correcting for detector cross-talk
using dedicated routines
\footnote{http://www.tng.iac.es/instruments/nics/files/crt\_nics7.f}
and then performing standard image preprocessing (flat fielding, bad pixels
and sky background corrections) in the IRAF environment. Individual images taken at a
fixed orientation were shifted and coadded.

\begin{figure}
\includegraphics[width=0.48\textwidth]{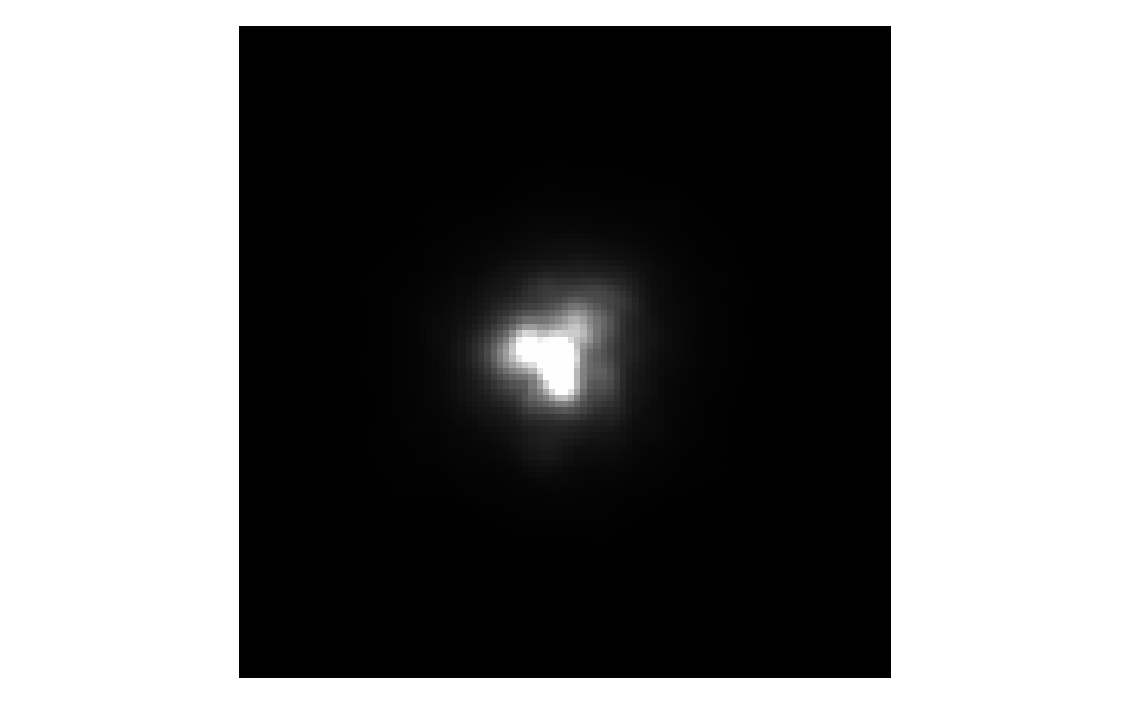}
\includegraphics[width=0.48\textwidth]{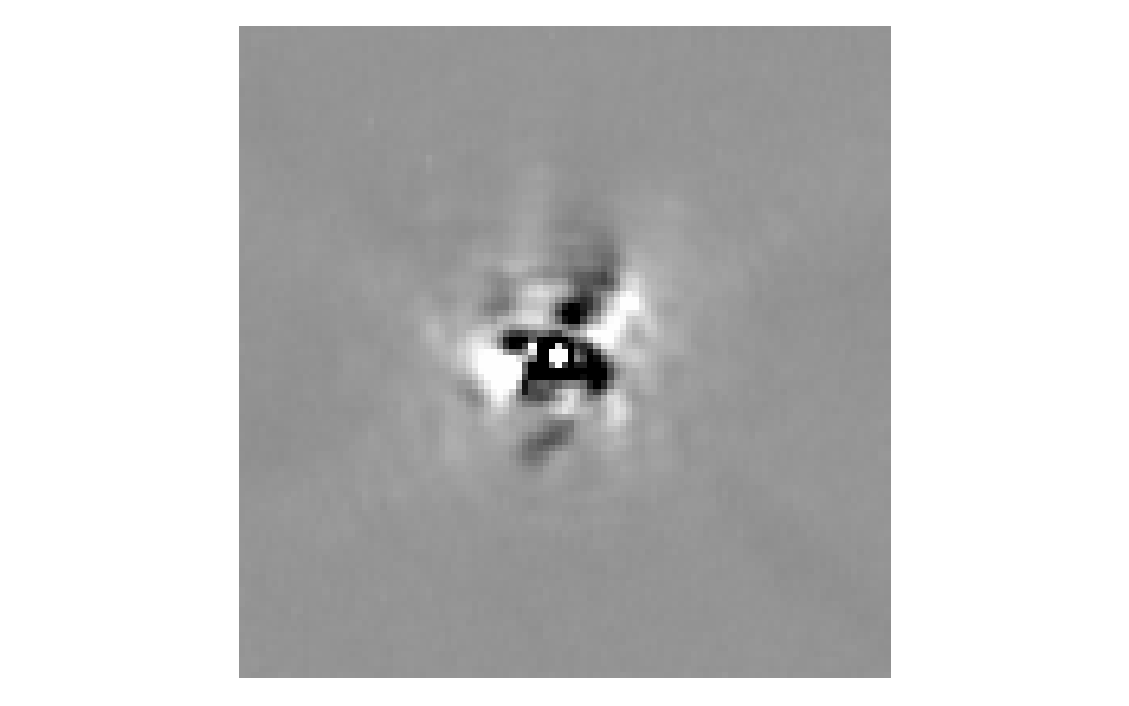}
\caption{
Images of \hd. Upper panel: image of \hd obtained with ADOPT@TNG.
A similar quasi-static speckle pattern can be seen.
Lower panel: difference between two images taken with two different field orientations, which
allows to improve significantly detection limits in the inner regions.
The field of view shown is 4.3x4.3\arcsec. The images are
displayed in linear gray scale.
}
\label{f:images}
\end{figure}

The successive analysis was optimized for the detection of companions in
different separation ranges. At small separations (from about 0.15 to 2 arcsec)
we selected the two best combined images taken at different field orientations on 2007 Oct 18.
They are shown in Fig.~\ref{f:images}. These two sets of images are
characterized by similar patterns of optical aberrations, and therefore,
considering their difference, most of the patterns cancel out in difference images
(Fig.~\ref{f:images}), improving significantly the detection limits (angular
differential imaging, \citet{2006ApJ...641..556M}).
In the differential
image, a true companion is expected to show two peaks, one positive and one
negative, at the same projected separation from the central star and position
angle displaced by 20$^\circ$ (the rotation angle between the two sets of images
in our case).
For detection at separations larger than about 2
arcsec, we summed all the images after an appropriate rotation,
obtaining a deep image over a field of about 10$\times$10 arcsec.

\begin{figure}
\resizebox{\hsize}{!}{
\includegraphics{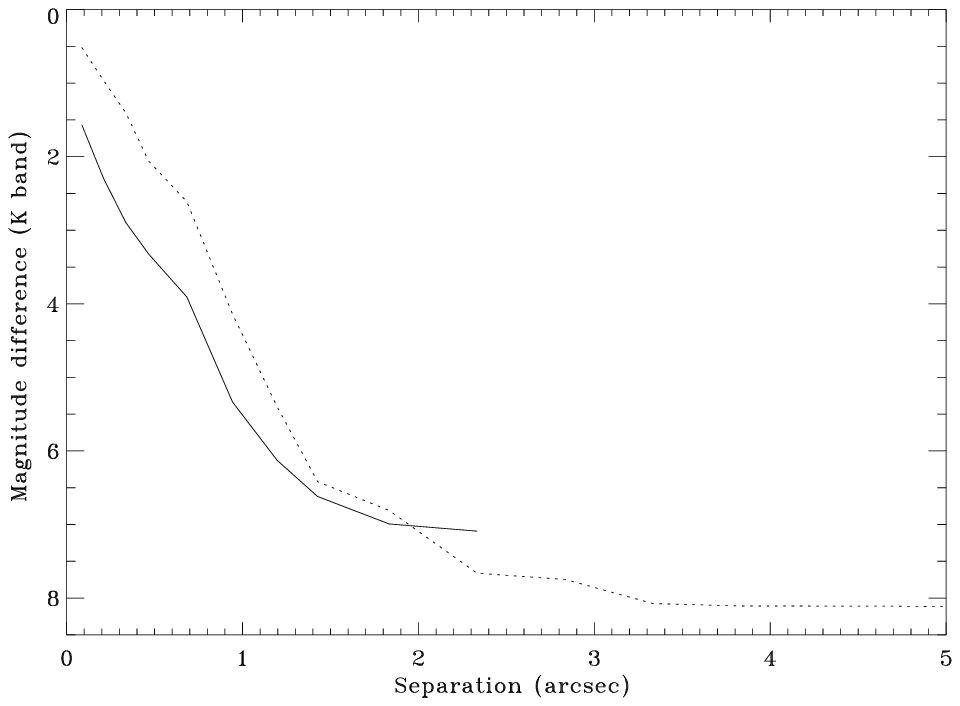}
}
\caption{
Detectability limits for companions of difference magnitude
around HD~17156 as a function of the projected separation in arcsec
Continuous line: limits on the difference image.
Dotted line: limits on the composite deep image.}
\label{f:maglim}
\end{figure}

No companion was seen in both the differential image at small
separation and in the deep combined image within 10 arcsec. The limit
for detection was fixed at peak intensities 5 times larger than
the dispersion over annuli at different radial separation. The results, both for the
differential image and the deep composite image are shown in Fig.~\ref{f:maglim}.

\begin{figure}
\resizebox{\hsize}{!}{
\includegraphics{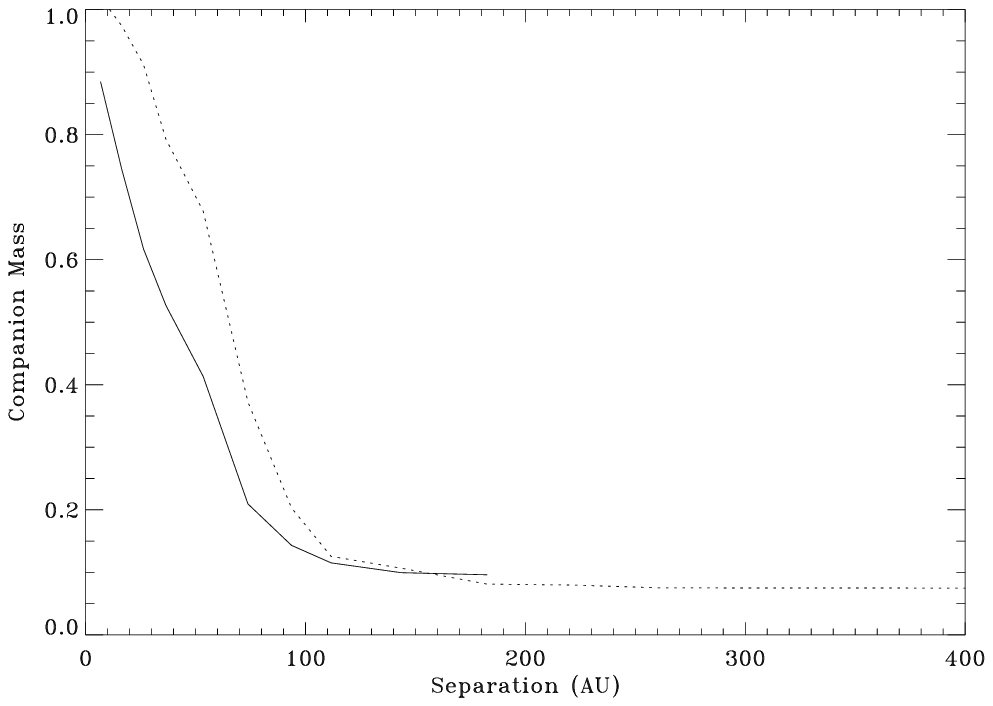}
}
\caption{
Detectability limits for stellar companions around HD 17156
as a function of the projected separation in AU.
Continuous line: limits on the difference image
dotted line: limits on the composite deep image.}
\label{f:masslim}
\end{figure}

The contrast limits derived above were transformed into limits on
companion masses using the mass-luminosity
relation by \cite{2000A&A...364..217D},
and projected separation in arcsec to AU using the
Hipparcos distance to the star (Fig.~\ref{f:masslim}).
A main-sequence companion can be
excluded at a projected separation between about 150 and 1000 AU
(the limit of image size).
At such separations only brown-dwarf or white-dwarf
companions are compatible with our detection limits.
At smaller separation, detectability worsens quickly, and only stars
with mass larger than about 0.4 M$_\odot$ can be excluded at projected
separations closer than $\sim50$ AU.

The residuals from the radial velocity orbital solution do not suggest
the occurrence of long term trends. This places further constraints on
the binarity of the target.  However, the timespan is rather short, and
the continuation of the radial velocity monitoring is mandatory for a
more complete view.

The available astrometric data from Hipparcos does not show evidence for
stellar companions either (no astrometric acceleration within the timespan
of the Hipparcos observations and no significant differences between

\section{Summary and Discussion}
\label{s:concl}
In this paper, we studied the characteristics of HD 17156 and its
transiting planets.
Stellar parameters (mass, radius, metallicity) agree quite well
with the previous study by F07.

Our measurement of the stellar radius of \hd obtained through the analysis
of the transit light-curve  is the same as the one obtained using  the
Stefan-Boltzmann law or the  Kervella calibration (Tab.\ref{t:pstar}). \cite{2008A&A...485..871G}
obtained a radius of 1.63 $\pm 0.2$
R$_\odot$, which is only marginally compatible  with our estimate 1.44 $\pm 0.08$ R$_\odot$.
On the one hand, to explain such a large stellar radius and the observed visual magnitude,
it would be necessary to add $\sim0.3$ mag of interstellar absorption, that at the
distance of  \hd is not realistic  (suggesting a mean  extinction of 4
mag/kpc) because \hd is located well inside of the Local Bubble, where
no strong  absorption is present, and the  maximum expected absorption
is few hundreds of mag.
On the other hand, also the comparison of the \cite{2008A&A...485..871G}
radius estimate with stellar models does not appear satisfactory:
it is not possible to find model  with a radius
that agrees with the observed  temperature  and metallicity  of \hd.
We conclude that the determination of the stellar radius, and by inference planetary radius,
\cite{2008A&A...485..871G} is overestimated by 15\%.

For  a  planet  of  3M$_{\rm J}$ and an age of $\sim2$~Gyr,  theoretical  models  of  planet
evolution \citep{2008A&A...482..315B} predict  a radius ranging between 0.9 and
1.1R$_{\rm J}$  as a function of  chemical  composition  of  the planet.  Our
determination of the radius of \hdb is $R_p =1.02  \pm 0.08$ R$_{\rm J}$.
This is in excellent agreement theoretical expectations.
Thus, the strong tidal heating effects on the planet do not appear to
contribute to significantly inflate its radius.

\cite{2007ApJ...659.1661F} suggested that \hdb,  due to its large orbital
eccentricity, can change its spectral type  during the orbit from a warm pL type
at apoastron to  a hotter pM type at  periastron. Their models suggest
that for pM class planets the observed radius at {\em R} band could be 5\% 
larger  than the  radius measured  at {\em B}  band (due  to  the increased
opacity of TiO and VO in  the {\em R} band). Comparing our radius measurements
in the {\em R}  band and  the measurements  obtained in the {\em B}  band
by \cite{2008A&A...485..871G}, we find identical results. This
result is however not significant because of the large error involved in radius
measurement. In order to obtain a significant difference of
the radius in {\em B} and {\em R}  band each measurement should be more
accurate than 0.03~R$_{\rm J}$.

Our  RV monitoring  of \hd  during  the 2007 December 3  transit does  not
confirm  the misalignment between  the stellar  spin and  planet orbit
axes claimed  by \cite{2008PASJ...60L...1N}, but it agrees instead
with the opposite finding by \cite{2008ApJ...683L..59C}.
We think our results are more robust because the other most accurate dataset
(HET in \citealt{2008ApJ...683L..59C}) does not cover the full transit,
leaving some uncertainties in their Rossiter modeling.
We then conclude that the projection on the sky of the stellar spin and
planet  orbit  axes are aligned to better than 10 deg.

Therefore  \hd  joins  most of the other  exoplanet
systems with available measurements of the Rossiter-McLaughlin effect in being compatible
with coplanarity. The only possible exception is represented by the
XO-3 system, for which \cite{2008A&A...488..763H} found
indications for a large departure
from coplanarity ($\sim 70^\circ$). However, as acknowledged by the authors,
this result should be taken as preliminary, because of the possibility
of unrecognized systematic errors for observations taken at large
airmass and with significant moonlight contamination.

Our result confirms that large deviations from coplanarity
between stellar spin and planet orbit  axes are at most rather rare.
Such a rarity had already been established
at a high confidence level for the ``classical''
Hot Jupiters in  short-period circular orbits.
For massive  eccentric  planets  the situation is less clear:
\object{HD  147506b} ($M=8.6$ M$_{\rm J}$ , $P=5.6$ days, $e=0.52$)  and \hd
($M=3.2$~M$_{\rm J}$ , $P=21$ days,
$e=0.67$) have projected inclinations below 10$^\circ$
while the possible detection of spin-orbit misalignment in the
XO-3 system ($M=12.5$~M$_{\rm J}$, $P=3.2$  days, $e=0.29$) still awaits
confirmation as discussed above.

The results of the spin-orbit alignment measurements for the \hd
system can be compared with the prediction of the planet scattering models.
A range of alignments can be the outcome of planet-planet scattering
\citep{2002Icar..156..570M}. Therefore, our indication for coplanarity does not exclude
planet-planet scattering in the HD~17156 system.
A larger number of transiting planets with significant
eccentricities have to be discovered and characterized to get
more conclusive inferences.

We also searched for stellar companions using adaptive optics,
to test the hypothesis of Kozai mechanism to explain the
large eccentricity of HD~17156~b.
We did not detect companions within 1\,000 AU, and our detection limits
allowed us to exclude main sequence companions with projected separations
from about 150 to 1\,000 AU.
This result makes unlikely the occurrence of a companion inducing
Kozai eccentricity oscillations on the planet, but this possibility
can not be yet completely rule out (companions at small projected
separation and faint white dwarfs and brown dwarfs companion still
possible). Continuation of radial velocity and photometric monitoring
will allow a more complete view on the existence of additional
companions at small separations.

\begin{acknowledgements}

This work was partially funded by PRIN 2006 "From disk to planetary systems:
understanding the origin and demographics of solar and extrasolar planetary
systems'' by INAF.
We thank the TNG director for time allocation in Director Discretionary Time.

\end{acknowledgements}

\end{document}